\documentclass[english,aps,prl, twocolumn]{revtex4-1}
\usepackage[T1]{fontenc}
\usepackage[latin9]{inputenc}
\usepackage{amsmath}
\usepackage{amssymb}
\usepackage{graphicx}
\usepackage{hyperref}

\makeatletter

\renewcommand\[{\begin{equation}}
\renewcommand\]{\end{equation}}
\newcommand{\xie}{\xi_\varepsilon}
\newcommand{\kF}{k_{\rm F}}
\newcommand{\Gfull}{\left\langle G(\omega) \right \rangle}
\newcommand{\Go}{G^{(0)}(\omega)}

\renewcommand{\vr}{{\mathbf{r}}}
\newcommand{\vk}{{\mathbf{k}}}

\usepackage{amssymb}
\usepackage{siunitx}
\usepackage{placeins}
\usepackage{braket}

\makeatother

\usepackage{babel}
\begin{document}

\title{Effects of nonmagnetic disorder on the energy of Yu-Shiba-Rusinov states}

\author{Thomas Kiendl}
\email{thomas.kiendl@fu-berlin.de}
\author{Felix von Oppen}
\author{Piet W. Brouwer}
\affiliation{Dahlem Center for Complex Quantum Systems and Fachbereich Physik, Freie Universit\"at Berlin, 14195, Berlin, Germany}

\begin{abstract}
 We study the sensitivity of Yu-Shiba-Rusinov states, bound states that form around magnetic scatterers in superconductors, to the presence of nonmagnetic disorder in both two and three dimensional systems. We formulate a scattering approach to this problem and reduce the effects of disorder to two contributions: disorder-induced normal reflection and a random phase of the amplitude for Andreev reflection. We find that both of these are small even for moderate amounts of disorder. In the dirty limit in which the disorder-induced mean free path is smaller than the superconducting coherence length, the variance of the energy of the Yu-Shiba-Rusinov state remains small in the ratio of the Fermi wavelength and the mean free path. This effect is more pronounced in three dimensions, where only impurities within a few Fermi wavelengths of the magnetic scatterer contribute. In two dimensions the energy variance is larger by a logarithmic factor because impurities contribute up to a distance of the order of the superconducting coherence length.
\end{abstract}
\maketitle

\section{Introduction}

Adding magnetic impurities to an $s$-wave superconductor induces bound states, whose excitation energy falls within the superconducting gap. This prediction goes back to works of Yu, Shiba, and Rusinov (YSR) in the 1960s \cite{YU1965,Shiba_1968,1969JETPL...9...85R}. In the meantime, numerous aspects of YSR states have been considered theoretically \cite{1969JETP...29.1101R, Zittartz_1972, Shiba_Cite_Hirschfeld_1988, Shiba_Cite_Balatsky_2006, Morr2006, PhysRevB.77.174516, Zitko2011, Yao2014, Hoffman2015, Kim2015, Meng2015, Zitko2016} and YSR states are now routinely observed by scanning tunneling spectroscopy on superconductors \cite{Yazdani_1997, Ji_2008, Ji_2010, Franke940, M_nard_2015, Ruby_2015, Hatter2015, Ruby_2016, choi_2016, 2017arXiv170103288K, Franke2017review}.

Interest in YSR states was recently renewed for several reasons. One reason is that experimental progress admits measurements of subgap spectra with much higher resolution than previously possible. This has triggered experimental and theoretical work exploring the basic properties of YSR states in more detail. It was recently found that reducing the dimensionality of the superconducting host from three to two
dimensions greatly increases the observed spatial extent of the YSR states, which is a consequence of the different power laws with which the YSR wavefunctions decay away from the impurity \cite{M_nard_2015, 2017arXiv170103288K}. Other recent work traced the origin of multiple YSR states to the crystal splitting of higher angular momentum channels \cite{Ruby_2015,Ruby_2016,Ji_2008,Hatter2015,choi_2016}.

Another reason is that chains of magnetic adatoms on superconductors have been proposed as a realization of a topological superconducting phase which harbors Majorana bound states at the ends of the chain, motivating several recent experimental studies of such systems \cite{yazdani2014, franke2015, meyer2016, feldman2017,  franke2017}. Majorana bound states are quasiparticles which are their own antiparticles and potential building blocks of a future topological quantum computer \cite{Kitaev20032,Nayak2008}. One way to think about this topological superconducting phase is in terms of an effective tight-binding model of hybridized YSR states \cite{Nadj_Perge_2013,  Pientka_2013, Klinovaja2013, Braunecker2013, Kim2014, Schecter2016}.

It is an important question to which degree YSR bound states are sensitive to potential (nonmagnetic) impurities in the superconductor. In the context of individual magnetic impurities, strong sensitivity to potential impurities  would make the YSR energies sample specific, reflecting the details of the impurity configuration in the vicinity of the magnetic atom. Similarly, topological superconductivity is known to be sensitive to disorder. Sensitivity of the YSR state to potential scatterers in the superconductor could thus be detrimental to the formation of a topological superconducting phase.

In this work, we characterize the sensitivity of YSR states in two and three dimensions to potential scatterers. We find that the YSR states are robust to disorder, even when the mean free path is shorter than the coherence length of the superconductor. The major condition for this robustness is that the disorder induced mean free path is large compared to the Fermi wavelength which is usually satisfied in conventional superconductors. Thus, our findings relax previous claims \cite{Hui_2015_2} that ultraclean superconductors are required for disorder to introduce only a small perturbation. These earlier results do not include a discussion of the Fermi wavelength, which we find to be a crucial parameter when considering the robustness of the YSR energy. We refer to appendix A for a more detailed discussion of the origin of the discrepancy with Ref. \cite{Hui_2015_2}.

This paper is structured as follows. In Sec.\ \ref{sec:Model} we introduce the model Hamiltonian on which our analysis is based. In Sec.\ \ref{sec:Perturbative-approach}, we review the YSR wavefunctions
in the absence of disorder and present a perturbative analysis of the effect of disorder on the YSR energies. Section \ref{sec:Scattering-approach} introduces a scattering approach, which in an approximate analytical approach, allows us to reduce the effects of disorder on the YSR energy to two contributions which can be discussed qualitatively based on symmetry arguments and a random walk model. In addition, we also employ the scattering approach for a numerical calculation beyond perturbation theory and compare it to the results obtained by perturbation theory. Finally, we conclude in Sec.\ \ref{sec:Conclusion}.

\section{Model\label{sec:Model}}

The system we consider is described by the Bogoliubov-de Gennes Hamiltonian
\begin{equation}
  H = \begin{pmatrix} H_0 + V_{\uparrow}(\vr) + U(\vr) & \Delta \\ \Delta & -H_0 - V_{\downarrow}(\vr) - U(\vr) \end{pmatrix},
  \label{eq:HBdG}
\end{equation}
where $H_0 = p^2/2m - \hbar^2 k_{\rm F}^2/2m$, with $m$ the (effective) mass and $k_{\rm F}$ the Fermi wavenumber, $\Delta$ the superconducting order parameter, which we choose to be real, $V_{\sigma}(\vr)$ the impurity potential, and $U(\vr)$ the disorder potential. 

Following Refs.\ \onlinecite{YU1965,Shiba_1968,1969JETPL...9...85R} we take the impurity to be a classical spin of magnitude $S$, located at the origin $\vr = 0$. Choosing the spin quantization axis along the impurity spin direction, the spin-dependent impurity potential has the form 
\begin{equation}
  V_{\sigma}(\vr) = (V_0 - J S \sigma )\delta_{\lambda}(\vr).
  \label{eq:H - impurity potential}
\end{equation}
Here, $\delta_{\lambda}(\vr)$ represents a short-ranged function with unit integral and range $\lambda \sim 1/k_{\rm F}$, $J$ is the exchange coupling strength, and $V_0$ is the strength of the potential scattering by the impurity.

For the disorder potential $U(\vr)$, we take a Gaussian white noise model, for which $U(\vr)$ has zero mean and variance
\begin{equation}
  \left\langle U(\mathbf{r})U(\mathbf{r}')\right\rangle =
  \frac{\hbar v_{\rm F}}{2 \pi \nu_0 \ell}
  \delta_{\lambda}\left(\mathbf{r}-\mathbf{r}'\right),
  \label{eq:correlator}
\end{equation}
where $\ell$ is the elastic mean free path, $v_{\rm F} = \hbar k_{\rm F}/m$ the Fermi velocity, and $\nu_0$ the density of states per spin direction. [In two dimensions ($d=2$) and three dimensions ($d=3$), one has $\nu_0 =  k_{\rm F}/2 \pi \hbar v_{\rm F}$ and $\nu_0 = k_{\rm F}^2/2\pi^2\hbar v_{\rm F}$, respectively.] For simplicity, we choose the same short-distance cutoff $\lambda$ for the disorder potential $U(\vr)$ and for the impurity potential $V_\sigma(\vr)$.

\begin{figure}
\includegraphics[width=1\columnwidth]{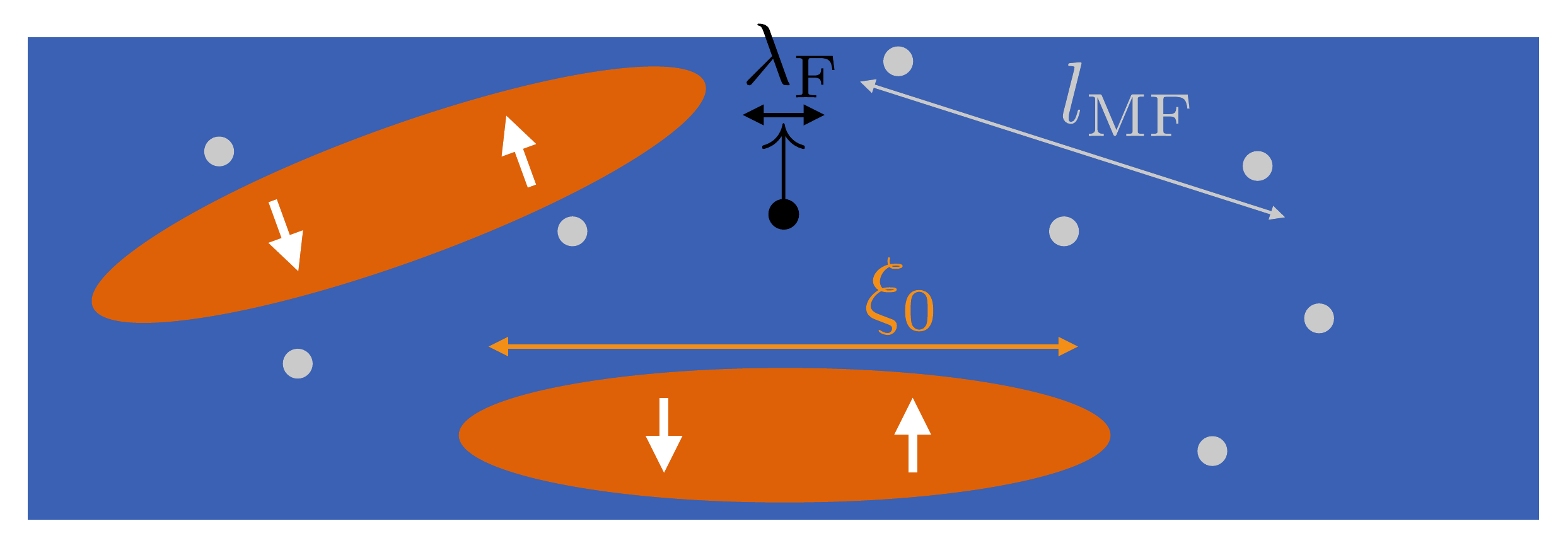}\caption{Sketch of the model used in this work and the relevant length scales.
A magnetic impurity with a size comparable to the Fermi wavelength
$\lambda_{F}$ is placed inside a superconductor. The coherence length
$\xi_{0}$ corresponds to the size of Cooper pairs (orange) that form
the superconducting ground state in the absence of disorder. Adding
nonmagnetic disorder (gray circles) introduces a mean free path $\ell$.
\label{fig:length scales}}
\end{figure}

The characteristic length scales of the system are illustrated in Fig. \ref{fig:length scales}. The superconductor is characterized by the clean-limit superconducting coherence length $\xi_0 = \hbar v_{\rm F}/\Delta$. For weak-coupling superconductors, one has $k_{\rm F} \xi_0 \gg 1$. Also, in superconductors that are good metals in the normal state, one has $k_{\rm F} \ell \gg 1$. We will assume that both inequalities are obeyed in the considerations that follow, but we will not make any assumptions concerning the relative magnitude of the mean free path $\ell$ and the coherence length $\xi_0$. Superconductors with $\ell \gg \xi_0$ are in the clean limit; superconductors with $\ell \ll \xi_0$ are in the dirty limit.

\section{Perturbative approach\label{sec:Perturbative-approach}}

In the presence of the impurity potential $V$, the Bogoliubov-de Gennes equation
\begin{equation}
  H \ket{\psi} = \varepsilon \ket{\psi},
  \label{eq:BdG}
\end{equation}
with $H$ given by Eq.\ (\ref{eq:HBdG}), has a bound-state solution with energy $|\varepsilon| < \Delta$. In this section, we review Rusinov's original calculation of the bound-state energy $\varepsilon_0$ for a clean superconductor \cite{1969JETPL...9...85R}. We then use this as a starting point to calculate the shift $\delta \varepsilon = \varepsilon - \varepsilon_0 $ to first order in the disorder potential $U$.

%We first calculate the disorder-induced energy shift of the 
%energy of the YSR state to lowest nontrivial order in the 
%impurity potential $V(\vr)$. 
%In the next Section we will show that the perturbative
%calculation is an excellent approximation for a three-
%dimensional superconductor ($d=3$) if $k_{\rm F} \ell \gg 1$, 
%whereas there are logarithmic corrections to the first-order 
%result in two dimensions ($d=2$).
We start with the general radially symmetric solution of the Bogoliubov-de Gennes equation (\ref{eq:BdG}) for $r > \lambda$, where $\lambda$ is the range of the potential $V_{\sigma}(\vr)$. For $k_{\rm F} \xi_0 \gg 1$, this reads
\begin{equation}
  \ket{\psi(\vr)} = 
  \frac{1}{\sqrt{\xi_{\varepsilon}}}
  \sum_{\pm} a_{\pm} \Phi_{\pm}(r) e^{-r/\xi_{\varepsilon}}
  \ket{\pm},
  \label{eq:general}
\end{equation}
where $a_{\pm}$ are complex coefficients, 
\begin{equation}
  \xi_{\varepsilon} = \frac{\hbar v_{\rm F}}{\sqrt{\Delta^2 - \varepsilon^2}}
\end{equation}
is the energy-dependent coherence length, $\Phi_{\pm}(r)$ solves the radial Schr\"odinger equation at $\varepsilon=0$ in the absence of superconductivity and takes different forms in two and three dimensions,
\begin{equation}
  \Phi_{\pm}(r) = 
  \left\{ \begin{array}{ll}
  \sqrt{\frac{k_{\rm F}}{4}} H_{0}^{\pm}(k_{\rm F} r) &
  d=2, \\
  \sqrt{\frac{k_{\rm F}^2}{4\pi}} h_0^{\pm}(k_{\rm F} r) &
  d=3,
  \end{array} \right.
\end{equation}
with $H^{\pm}$ and $h^{\pm}$ the Hankel and spherical Hankel functions, respectively, and $\ket{\pm}$ are two-component spinors,
\begin{equation}
  \ket{\pm} = \frac{1}{\sqrt{2}}
  \begin{pmatrix}1\\
  e^{\mp i\eta(\varepsilon)}
  \end{pmatrix},
  \label{eq:particle-hole spinor in the clean case}
\end{equation}
with the Andreev phase 
\begin{equation}
  \eta(\varepsilon)=\arccos(\varepsilon/\Delta)\label{eq:Andreev phase in the clean case}.
\end{equation}

To determine the coefficients $a_{\pm}$ we use the boundary condition at $r = \lambda$ imposed by the magnetic impurity. Following Rusinov's original derivation \cite{1969JETPL...9...85R}, we formulate this boundary condition in terms of scattering phase shifts $\phi_{\sigma}$ for electrons with spin $\sigma$. We may neglect the effect of superconductivity on the phase shifts $\phi_{\sigma}$ since $\lambda \ll \xi_0$. The scattering phases $\phi_{\sigma}$ can be related to the impurity potential (\ref{eq:H - impurity potential}) by \cite{1969JETPL...9...85R,Clogston_1962, footnote1}
\begin{equation}
  \tan \phi_{\sigma} =-\pi\nu_{0}(V_{0}-\sigma JS).
  \label{eq:phase - potential - tan(phi)}
\end{equation}

We note that in the general solution (\ref{eq:general}) the upper component multiplied by $a_+$ ($a_-$) describes a radial wave for an electron with spin up moving away from (towards) the origin. Hence, 
\begin{equation}
  a_+ = e^{2 i \phi_{\uparrow}} a_-.
  \label{eq:arel1}
\end{equation}
Similarly, the lower component multiplied by $a_+$ ($a_-$) describes a radial wave moving towards (away from) the origin for a hole in the spin-down band. Taking into account the phase factors $e^{\pm i \eta(\varepsilon)}$ in the lower component of the spinor (\ref{eq:particle-hole spinor in the clean case}), we find the relation
\begin{equation}
  a_- e^{i \eta(\varepsilon)} =
  e^{-2 i \phi_{\downarrow}} a_+ e^{-i \eta(\varepsilon)}.
  \label{eq:arel2}
\end{equation}
Combining these two equations we find the YSR-state energy
\begin{equation}
  \varepsilon_0 = \pm \Delta \cos (\phi_{\uparrow} - \phi_{\downarrow})
  \label{eq:E0}
\end{equation}
for a magnetic impurity in an (otherwise) clean superconductor. The solution (\ref{eq:general}) is properly normalized if $|a_+| = |a_-| = 1$ (up to corrections that are small in the limit $k_{\rm F} \xi_0$, $\xi_0/\lambda \gg 1$). Note that $|\varepsilon_0| < \Delta$ if and only if $\phi_{\uparrow} \neq \phi_{\downarrow})$, {\em i.e.}, if the impurity is magnetic.

We now calculate the change $\delta \varepsilon$ of the energy of the YSR state to first order in the disorder potential $U(\vr)$. We assume that $U(\vr) = 0$ for $r < \lambda$, {\em i.e.}, the disorder potential does not overlap with the potential of the magnetic impurity. To first order in $U(\vr)$ the change $\delta \varepsilon$ is
\begin{equation}
  \delta \varepsilon = \int\mathrm{d}\mathbf{r}\bra{\psi(\vr)}
  U 
  (\mathbf{r})\tau_{z}\ket{\psi(\vr)},
\end{equation}
where $\tau_z$ is the Pauli matrix in particle-hole space and the spinor wavefunction $\ket{\psi(\vr)}$ is given by the general solution (\ref{eq:general}), with the coefficients $a_{\pm}$ satisfying the relations (\ref{eq:arel1}) and (\ref{eq:arel2}). Using the relation $\Phi_+(k_{\rm F} r) = [\Phi_-(k_{\rm F} r)]^*$ this can be simplified as
\begin{align}
  \delta \varepsilon =& \sin \eta(\varepsilon_{0})
  \int\mathrm{d}\vr U(\vr)
  e^{-2 r/\xi_{\varepsilon_0}}
  \nonumber \\ &\, \mbox{} \times
  \mathrm{Im}\, e^{-i\eta(\varepsilon_{0}) + 2 i \phi_{\uparrow}}
  [\Phi_{+}(r)]^{2}.
  \label{eq:deltaE_explicit}
\end{align}
Using the correlator (\ref{eq:correlator}), we obtain the variance
\begin{align}
  \langle \delta \varepsilon^2 \rangle =&\,
  \frac{2 \hbar v_{\rm F}}{\pi \nu_0 \ell \xi_{\epsilon_0}^2}  
  \sin^{2}\eta(\varepsilon_{0})
  \label{eq:pert-dE-ddim formula}
  \\ &\, \mbox{} \times
  \int d\vr \left\{\mathrm{Im}\, e^{-i\eta(\varepsilon_{0}) + 2 i \phi_{\uparrow}}
  [\Phi_{+}(r)]^{2}\right\}^{2} e^{-4 r/\xi_{\varepsilon_{0}}}.
\nonumber 
\end{align}

In two dimensions the main contribution to the integral (\ref{eq:pert-dE-ddim formula}) comes from $r \sim \xi_{\varepsilon_0}$. The integral is convergent at the lower limit $r \downarrow 0$, so that the short-distance cutoff $\lambda$ can be taken to zero. We then find
\begin{equation}
  \langle \delta \varepsilon^2 \rangle =
  \frac{\Delta^2 (\xi_0/\xi_{\varepsilon_0})^4}{\pi k_{\rm F} \ell } \log (k_{\rm F} \xi_{\varepsilon_0}).
  \label{eq:2D - 1st order result}
\end{equation}
In three dimensions the integral (\ref{eq:pert-dE-ddim formula}) is dominated by short distances $r \sim \lambda \sim k_{\rm F}^{-1}$ and the short-distance cutoff $\lambda$ is needed to ensure convergence. In this case we find
\begin{equation}
  \langle \delta \varepsilon^2 \rangle \sim
  \frac{\Delta^2 (\xi_0/\xi_{\varepsilon_0})^4}{k_{\rm F}\ell},
  \label{eq:ballistic - 3D integral}
\end{equation}
where a numerical prefactor depends on the precise way in which the short-distance regularization is implemented. Note that in three dimensions and with $\varepsilon_0$ well inside the gap such that $\xi_0/\xi_{\varepsilon_0} \sim 1$, the variance $\langle \delta \varepsilon^2 \rangle$ does not depend on $k_{\rm F} \xi_{\varepsilon_0}$.

In two dimensions the root-mean-square fluctuations are parametrically smaller than the superconducting gap $\Delta$ if the condition $k_{\rm F} \ell \gg \log(k_{\rm F} \xi_{0})$ is met. This condition only weakly depends on the superconducting coherence length $\xi_0$. In three dimensions the condition is $k_{\rm F} \ell \gg 1$, which is independent of $\xi_0$. The latter condition $k_{\rm F} \ell \gg 1$ is typically met in superconductors that are good metals in the normal state, such as Pb or Al. 

\section{Scattering approach\label{sec:Scattering-approach}}

In this section, we present a numerical calculation of the YSR-state energies that takes higher-order contributions in the disorder potential $U(\vr)$ into account. The calculation makes use of a relation between the YSR-state energy $\varepsilon$ and the scattering matrix $S(\varepsilon)$ of the superconductor for radial waves moving towards and from the origin $\vr = 0$. We first describe this relation and the calculation of $S(\varepsilon)$ separately, and then proceed with a calculation of the variance $\langle \delta \varepsilon^2 \rangle$.

\subsection{Relation between YSR-state energy and scattering amplitudes}

To define the scattering matrix $S$  we introduce a narrow shell $r_0 < r < r_0 + \delta r$ around the impurity in which the superconducting order parameter $\Delta$ as well as the potentials $V$ and $U$ are set to zero. (At the end of the calculation, we will send the shell width $\delta r \to 0$.) We choose $r_0 \sim \lambda \lesssim 1/k_{\rm F}$. The solution of the Bogoliubov-de Gennes equation may be assumed to be radially symmetric, so that it can be expanded as
\begin{equation}
  \ket{\psi(\vr)} = \sum_{\pm} 
  \left[ a_{{\rm e},\pm} \ket{\Phi_{{\rm e},\pm}(\vr)}
  + a_{{\rm h},\pm} \ket{\Phi_{{\rm h},\pm}(\vr)} \right],
  \label{eq:expansion0}
\end{equation}
where $\ket{\Phi_{{\rm e},\pm}}$ and $\ket{\Phi_{{\rm e},\pm}}$ represent flux-normalized electron-like (e) and hole-like (h) waves propagating radially outward ($+$) or inward ($-$). In two dimensions one has
\begin{align}
  \ket{\Phi_{{\rm e},\pm}(\vr)} &= 
  \sqrt{\frac{k_{\rm F}}{4 v_{{\rm F}}}}
  H_{0}^{\pm}(k_{\rm F} r) 
  \begin{pmatrix} 1 \\ 0 \end{pmatrix},\nonumber \\
  \ket{\Phi_{{\rm h},\pm}(\vr)} &= 
  \sqrt{\frac{k_{\rm F}}{4 v_{{\rm F}}}}
  H_{0}^{\mp}(k_{\rm F} r) 
  \begin{pmatrix} 0 \\ 1 \end{pmatrix},
\end{align}
whereas in three dimensions
\begin{align}
  \ket{\Phi_{{\rm e},\pm}(\vr)} &= 
  \sqrt{\frac{k_{\rm F}^2}{4 \pi v_{{\rm F}}}}
  h_{0}^{\pm}(k_{\rm F} r) 
  \begin{pmatrix} 1 \\ 0 \end{pmatrix}, \nonumber \\
  \ket{\Phi_{{\rm h},\pm}(\vr)} &=
  \sqrt{\frac{k_{\rm F}^2}{4 \pi v_{{\rm F}}}}
  h_{0}^{\mp}(k_{\rm F} r) 
  \begin{pmatrix} 0 \\ 1 \end{pmatrix}.
  \label{eq:modes at the origin}
\end{align}
The solution of the Bogoliubov-de Gennes equation for $r > r_0 + \delta r$ yields two linear relations for the four amplitudes $a_{{\rm e},\pm}$ and $a_{{\rm h},\pm}$, which have the general form
\begin{equation}
  \begin{pmatrix} a_{{\rm e},-} \\ a_{{\rm h},-} \end{pmatrix}
  =
  S(\varepsilon) \begin{pmatrix} a_{{\rm e},+} \\ a_{{\rm h},+} \end{pmatrix}, \ \
  S = \begin{pmatrix} r_{\text{ee}} & r_{\text{eh}}\\
  r_{\text{he}} & r_{\text{hh}}
\end{pmatrix}.
  \label{eq:a1}
\end{equation}
The matrix $S$ is the scattering matrix of the superconductor for radial waves around the origin $\vr = 0$. The coefficients $r_{\rm ee}$ and $r_{\rm hh}$ are the amplitudes for normal reflection of electrons and holes, respectively, whereas $r_{\rm he}$ and $r_{\rm eh}$ describe Andreev reflection of electrons into holes and vice versa. Time-reversal symmetry 
%(which is present in the Bogoliubov-de Gennes Hamiltonian (\ref{eq:HBdG}) for $r > \lambda$) 
and particle-hole symmetry enforce the constraints 
\begin{equation}
  S(\varepsilon) = S^{\rm T}(\varepsilon) = \tau_y S(-\varepsilon)^* \tau_y,
\end{equation}
where $\tau_y$ is a Pauli matrix in particle-hole space.

In the absence of the disorder potential $U(\vr)$, one has $r_{\rm ee} = r_{\rm hh} = 0$, and $r_{\rm eh} = r_{\rm he} = e^{-i \eta(\varepsilon)}$, with $\eta(\varepsilon)$ defined in Eq.\ (\ref{eq:Andreev phase in the clean case}). This reproduces Eq. (\ref{eq:general}). In the presence of the disorder potential $U(\vr)$, all four coefficients are in general nonzero. As we will show below, the difference with the clean case is small when $k_{\rm F} \ell \gg 1$, so that we may write
\begin{equation}
  S(\varepsilon) = e^{-i \eta(\varepsilon)} \tau_x
%  \begin{pmatrix} 0 & e^{-i \eta(\varepsilon)}
% \\ e^{-i \eta(\varepsilon)} & 0 \end{pmatrix}
  \left(\openone - i \delta A(\varepsilon) \right),
  \label{eq:full S matrix}
\end{equation}
where $\delta A(\varepsilon)$ is small. The conditions that $S(\varepsilon)$ be unitary, symmetric, and particle-hole symmetric imply 
\begin{equation}
  \delta A(\varepsilon) = \delta A(\varepsilon)^{\dagger} = \tau_x \delta A(\varepsilon)^{\rm T} \tau_x = -\tau_z \delta A(-\varepsilon) \tau_z.
\end{equation}
We therefore parameterize
\begin{equation}
  \delta A(\varepsilon) = 
  \begin{pmatrix}
  \delta \eta(\varepsilon) & \delta r(\varepsilon) \\
  \delta r(\varepsilon)^* & \delta \eta(\varepsilon)
  \end{pmatrix},
  \label{eq:dA definition}
\end{equation}
where $\delta r(\varepsilon)$ is a complex, symmetric function of energy $\varepsilon$, which represents disorder-induced normal reflection, whereas $\delta \eta(\varepsilon)$ is a real, antisymmetric function of $\varepsilon$, which represents a disorder-induced shift of the Andreev reflection phase $\eta(\varepsilon)$. 

As discussed above, the solution of the Bogoliubov-de Gennes equation for $r < r_0$ yields two additional relations between the amplitudes $a_{{\rm e},\pm}$ and $a_{{\rm h},\pm}$,
\begin{equation}
  a_{{\rm e},+} = e^{2 i \phi_{\uparrow}} a_{{\rm e},-}, \ \
  a_{{\rm h},-} = e^{-2 i \phi_{\downarrow}} a_{{\rm h},+}.
  \label{eq:a2}
\end{equation}
Note that in Eq. (\ref{eq:a2}), we assumed that the size of the magnetic impurity is small compared to the coherence length, corresponding to $\kF \xi_0  \gg 1$. By taking this limit we can neglect scattering from electrons to holes at the position of the impurity.

A nontrivial solution of Eqs.\ (\ref{eq:a1}) and (\ref{eq:a2}) exists if
\begin{equation}
  \det\left[ \begin{pmatrix} e^{2 i \phi_{\uparrow}} & 0 \\
  0 & e^{-2 i \phi_{\downarrow}} \end{pmatrix}
  S(\varepsilon) - \begin{pmatrix} 1 & 0 \\ 0 & 1 \end{pmatrix} \right] = 0.
\end{equation}
This gives
\begin{align}
  \varepsilon =& \varepsilon_0 + \delta \varepsilon,\nonumber\\
  \delta \varepsilon =& \frac{\xi_0 \Delta}{\xi_{\varepsilon_{0}}}
  \left[ \delta \eta(\varepsilon_{0})
  + \mbox{Re}\, e^{-i (\phi_{\uparrow} + \phi_{\downarrow})}
  \delta r(\varepsilon_0) \right],
  \label{eq:der}
\end{align}
to lowest order in $\delta A$.
This equation is central to our analysis, since it allows us to calculate the energy shift $\delta \varepsilon$ from the scattering matrix $S$.

\subsection{Qualitative discussion}

Next, we employ a semiclassical picture to qualitatively discuss why both $\delta \eta$ and $\delta r$ are expected to be small. Within this semiclassical picture, Andreev reflection is retroreflection, {\em i.e.}, after Andreev reflection a hole retraces the path of the incident electron (or vice versa). Because the phases of electron and hole wavefunctions are correlated, see, {\em e.g.}, Eq.\ (\ref{eq:general}), no net phase is accumulated, with the exception of the Andreev phase $\eta(\varepsilon)$. If $k_{\rm F} \ell \gg 1$ this semiclassical picture remains valid in the presence of a disorder potential $U(\vr)$. This explains why $\delta \eta$, corresponding to a shift of the Andreev phase, is small if $k_{\rm F} \ell \gg 1$.

In fact, since $\delta \eta(\varepsilon)$ is an antisymmetric function of $\varepsilon$, one must have $\delta \eta(0) = 0$, so that there is no contribution to the YSR-state energy shift from the phase shift $\delta \eta$ for YSR states with energy near the center of the superconducting gap. Instead, for small YSR-state energies, the residual disorder-induced fluctuations are dominated by the normal reflection amplitude $\delta r$. An estimate of the size of the YSR-energy fluctuations can be obtained by estimating $\delta r$ as the amplitude that a particle returns to the origin $\vr = 0$ (up to a distance $1/k_{\rm F}$) within a time $\xi_{\varepsilon}/v_{\rm F}$. In the two-dimensional case we then find for the dirty superconductor limit $\xi_0 \gg \ell$
\begin{align}
  |\delta r(\varepsilon)|^2 \sim&\, \int_{\lambda/v_{\rm F}}^{\ell/v_{\rm F}}
  dt \frac{1}{k_{\rm F} \ell t}
  +
  \int_{\ell/v_{\rm F}}^{\xi_\varepsilon/v_{\rm F}}
  dt \frac{2}{k_{\rm F} \ell t}
  \nonumber \\ \sim&\,
  \frac{1}{k_{\rm F} \ell } \log(k_{\rm F}  \xi_{\varepsilon}^2/\ell).
  \label{eq:2D-prediction-multipleScattering}
\end{align}
The first integral covers ballistic propagation times $t \lesssim \tau$ and the second diffusive times $t \gtrsim \tau$, where $\tau = \ell/v_{\rm F}$ is the elastic mean free time. The integrands give the return probabilities per unit time, which, in the diffusive regime, is multiplied by a factor two due to coherent backscattering. In the second line the short-distance cutoff $\lambda$ was replaced by $1/k_{\rm F}$. In the ultraclean limit $\xi_0 \ll \ell$ the second term in Eq. (30) is absent and the upper integration limit in the first is $\xi_\varepsilon$, which gives 
\begin{equation}
	|\delta r(\varepsilon)|^2 \sim \frac{1}{k_{\rm F} \ell } \log(k_{\rm F}  \xi_{\varepsilon})
\end{equation}
In three dimensions, the return probability is dominated by the ballistic regime $t \lesssim \tau$ and one finds
\begin{equation}
  |\delta r(\varepsilon)|^2 \sim 
  \int_{\lambda/v_{\rm F}}^{\tau}
  dt \frac{1}{k_{\rm F}^2 \ell v_{\rm F} t^2}
%  \nonumber \\ \sim&\,
  \sim \frac{1}{k_{\rm F} \ell}.
\end{equation}
In three dimensions, the estimate is consistent with the smallness of the first-order perturbation theory results of Sec.\ \ref{sec:Perturbative-approach}. In two dimensions and in the dirty limit, multiple scattering changes the argument of the logarithm in Eq.~(\ref{eq:2D-prediction-multipleScattering}) by a factor $\xi_\epsilon/\ell$ compared to the perturbative result~(\ref{eq:2D - 1st order result}).

\subsection{Numerical calculation of the scattering matrix}

\begin{figure}
\includegraphics[width=1\columnwidth]{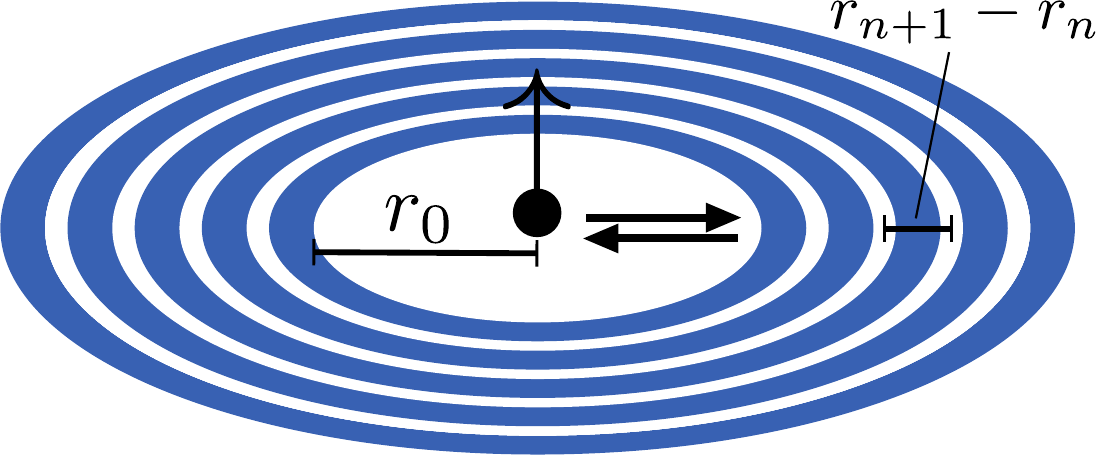}
\caption{
Sketch of the numerical method in two dimensions. The disordered superconductor is cut into thin circular slices (blue), with infinitesimally thin clean, metallic shells (white) in between. The scattering matrices of the slices are calculated in the narrow limit and combined into the full scattering matrix $S(\varepsilon)$, which describes the reflection of spherical symmetric waves (black) that are propagating from the magnetic impurity into the bulk.
\label{fig:slicing sketch}}
\end{figure}

Our strategy for an efficient numerical calculation can be outlined as follows. First, we slice the superconductor into thin circular ($d=2$) or spherical ($d=3$) pieces, as illustrated in Fig. (\ref{fig:slicing sketch}), and calculate the scattering matrix for each piece. Next, we add the pieces together by concatenating their scattering matrices to obtain the total scattering matrix $S(\varepsilon)$. Finally, using Eqs. (\ref{eq:full S matrix}) and (\ref{eq:der}) we can relate this scattering matrix to the energy of the YSR state.

Note that within this approach, we need to include non-zero angular momentum channels due to two reasons. First, disorder breaks rotational symmetry thus mixing different angular momentum modes. Second, the modes defined in Eq. (\ref{eq:modes at the origin}) describe all propagation modes at radii $k_{\rm{F}} r \lesssim 1$. There, non-zero angular momentum modes are evanescent. This can be seen by considering the spherical representation of the momentum operator, which results in a repulsive potential that diverges $\sim 1/r^2$ at the origin for all modes but the zero angular momentum one. However, when going to larger radii non-zero angular momentum modes start to propagate and thus cannot be excluded.

Our approach starts, by formally inserting a sequence of $N+1$ infinitesimally thin ideal shells into the disordered superconductor at radii $r_n$, $n=0,1,\ldots,N$. Within these shells the superconducting order parameter $\Delta$ and the impurity potential $U(\vr)$ are zero. The distance between the shells is $r_{n+1}-r_n \ll \ell$. In each shell the wavefunction can be expanded as [compare with Eq.\ (\ref{eq:expansion0})]
\begin{equation}
  \ket{\psi(\vr)} = \sum_{\nu} a^{(n)}_{\nu} \ket{\Phi_{\nu}(\vr)}, 
  \label{eq:expansion}
\end{equation}
where $\nu$ is a composite index representing particle/hole degrees of freedom (e or h), propagation direction (radially outward, $+$, or inward, $-$, and the angular momentum label $m$ (in two dimensions) or the angular momentum labels $l$, $m$ (in three dimensions). 

The flux-normalized basis functions $\ket{\Phi_{\nu}}$ are
\begin{align}
  \ket{\Phi_{{\rm e},m \pm}(\vr)} &=
   \sqrt{\frac{k_{\rm F}}{4 v_{{\rm F}}}}
  H_{m}^{\pm}(k_{\rm F} r) e^{i m \varphi}
  \begin{pmatrix} 1 \\ 0 \end{pmatrix}, \nonumber\\
  \ket{\Phi_{{\rm h},m \pm}(\vr)} &=
   \sqrt{\frac{k_{\rm F}}{4 v_{{\rm F}}}}
  H_{m}^{\mp}(k_{\rm F} r) e^{i m \varphi}
  \begin{pmatrix} 0 \\ 1 \end{pmatrix}
  \label{eq:2D - modes}
\end{align}
in two dimensions and
\begin{align}
  \ket{\Phi_{{\rm e},lm \pm}(\vr)} &= 
  \sqrt{\frac{k_{\rm F}^2}{4 \pi v_{{\rm F}}}}
  h_{l}^{\pm}(k_{\rm F} r) Y_{lm}(\theta,\varphi)
  \begin{pmatrix} 1 \\ 0 \end{pmatrix}, \nonumber \\
  \ket{\Phi_{{\rm h},lm \pm}(\vr)} &= 
  \sqrt{\frac{k_{\rm F}^2}{4 \pi v_{{\rm F}}}}
  h_{l}^{\mp}(k_{\rm F} r) Y_{lm}(\theta,\varphi)
  \begin{pmatrix} 0 \\ 1 \end{pmatrix}
\end{align}
in three dimensions, where the $Y_{lm}(\theta,\phi)$ are spherical harmonics and the $H_m$ ($h^{\pm}_l$) are (spherical) Hankel functions. The solution of the Bogoliubov-de Gennes equation for $r_n < r < r_{n+1}$ gives a linear relationship between the coefficients $a^{(n)}$ and $a^{(n+1)}$, which has the form (vector notation is implied for all indices not listed explicitly)
\begin{equation}
  \begin{pmatrix} a^{(n)}_{-} \\ a^{(n+1)}_{+} 
  \end{pmatrix}
  =
  {\cal S}^{(n)}(\varepsilon)
  \begin{pmatrix} a^{(n+1)}_{-} \\
  a^{(n)}_{+} 
  \end{pmatrix}  
  \label{eq:an}
\end{equation}
where ${\cal S}^{(n)}(\varepsilon)$ is the scattering matrix between the shells at $r_n$ and $r_{n+1}$.

If $r_{n+1}-r_{n}$ is much smaller than the mean free path $\ell$, the scattering matrix ${\cal S}^{(n)}(\varepsilon)$ can be calculated using the Born approximation, 
\begin{equation}
  {\cal S}^{(n)} = [\openone - i {\cal T}^{(n)}/2][\openone + i {\cal T}^{(n)}/2]^{-1},\label{eq:numerical Sn}
\end{equation}
with
\begin{equation}
  {\cal T}^{(n)}_{\nu'\nu} =
  \int_{r_n < r < r_{n+1}} d\vr
  \bra{\psi_{\nu'}(\vr)} H'_{\varepsilon}\ket{\psi_{\nu}(\vr)},\label{eq:Tmatrix definition}
\end{equation}
with $H'_{\varepsilon} = \Delta \tau_x + U(\vr) \tau_z - \varepsilon$. We refer to the appendix for explicit expressions for the matrices ${\cal T}^{(n)}$.  

To truncate the hierarchy of Eqs. (\ref{eq:an}) we set the disorder potential $U(\vr)$ to zero for $r > r_N$, which gives the relation
\begin{equation}
  a^{(N)}_- = e^{-i \eta(\varepsilon)} \tau_x a^{(N)}_+.
  \label{eq:aN}
\end{equation}
Further, for nonzero angular momentum indices $m$ or $l$ the Hankel functions $H_{m}$ and $h_{l}$ diverge for $k_{\rm F} r \lesssim \pi|m|/2$, $k_{\rm F} r \lesssim \pi l/2$, respectively. In that case regularity of  $\ket{\psi(\vr)}$ imposes that the corresponding coefficients $a_+^{(n)}$ and $a_-^{(n)}$ must be equal. In particular, 
%since the condition $k_{\rm F} r \lesssim \pi|m|/2$, $\pi l/2$ is met for all nonzero $m$ and $l$ at $r = r_0$, 
we have
\begin{equation}
  a_{{\rm e},+(l)m}^{(0)} = a_{{\rm e},-(l)m}^{(0)},\ \
  a_{{\rm h},+(l)m}^{(0)} = a_{{\rm h},-(l)m}^{(0)}
  \label{eq:a0}
\end{equation}
for all $m \neq 0$ ($d=2$) or $l>0$ ($d=3$). Similarly, this observation allows us to truncate the sum over modes $l$ and $m$ to the number of propagating angular momentum modes at the largest distance required for the calculation of $S(\varepsilon)$, which is $r \sim \xi_\varepsilon$.

Combining Eqs.\ (\ref{eq:an}), (\ref{eq:aN}), and (\ref{eq:a0}) we can eliminate all amplitudes $a^{(n)}$ with $n \ge 1$ and calculate the scattering matrix $S(\varepsilon)$ describing the (Andreev) reflection of radially outgoing waves at the origin. The procedure becomes numerically exact in the limit $r_{n+1} - r_{n} \to 0$, $r_N \to \infty$. In practice, to achieve convergence it is sufficient that $r_{n+1} - r_{n} \lesssim \ell $ and if $r_N \sim \xi_{\varepsilon}$ because of the exponential decay of the wavefunction in the superconductor. In the numerical simulations, the short-distance cutoff is fixed to $r_0 = 0$ in two and $r_0 = 1/\kF$ in three dimensions.

\begin{figure}
\includegraphics[width=1\columnwidth]{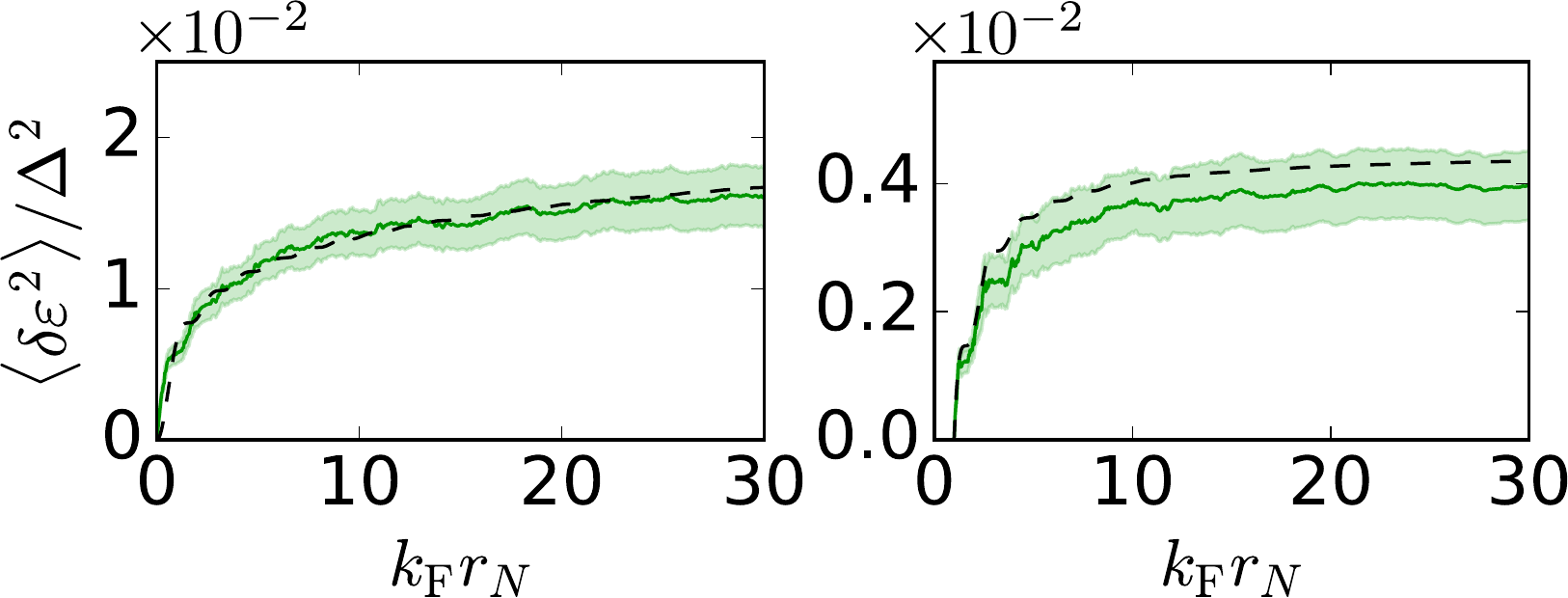}\caption{\label{fig:convergence} YSR energy shift (solid lines) versus disorder cutoff length $r_N$ in two dimensions (left panel) and three dimensions (right panel), obtained from the numerical scattering approach. We choose $\xi/\ell = 10$ with $k_{\rm F} \ell = 100$. When including the cutoff $r_N$, our perturbative approach (dashed lines) fits well with the numerical results. The shaded region shows the numerical standard error.}
\end{figure}

Figure \ref{fig:convergence} shows examples of the convergence behaviors in two and three dimensions. In two dimensions the numerical scattering matrix calculation converges slowly, in agreement with results from perturbation theory which predicts a logarithmic convergence at a length scale of the order of the coherence length. In contrast, our numerical results for three dimensional systems converge after a few Fermi wavelengths $2 \pi/k_{\rm F}$ and also agree well with our results derived by perturbation theory.

\begin{figure}
\noindent\begin{minipage}[t]{1\columnwidth}%
\includegraphics[width=1\columnwidth]{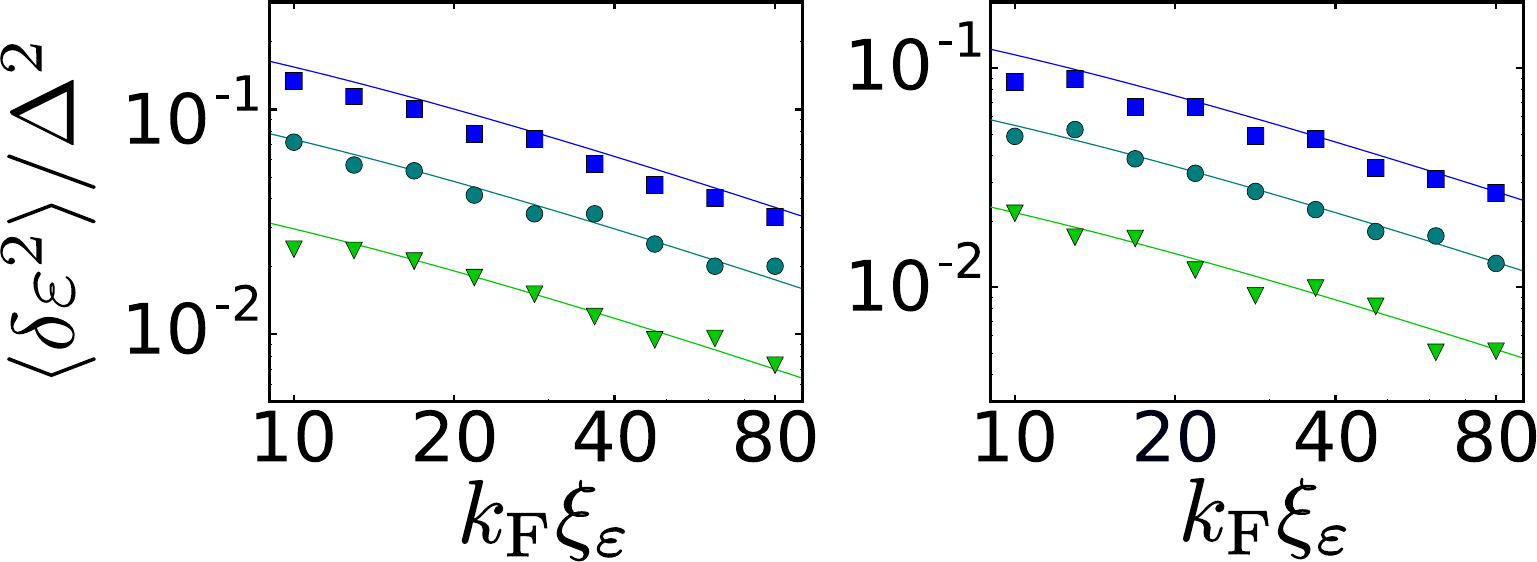}%
\end{minipage}\caption{YSR energy variance versus $k_{\rm F} \xi_\varepsilon$ in two dimensions. The data points are for $\xi_\varepsilon/\ell=0.5$ (triangles), $1$ (circles) and $2$ (squares). The scattering phase shifts $\phi_{\uparrow} = - \phi_{\downarrow}$ are chosen such that YSR energy in the absence of disorder is $\varepsilon/\Delta = 0$ (left) and $\varepsilon/\Delta = 0.37$ (right). The solid lines show the lowest-order perturbation theory result from Eq. (\ref{eq:2D - 1st order result}). Error bars are of the size of the markers. \label{fig:2D Energy-variance-over-kF}}
\end{figure}

In Fig.~\ref{fig:2D Energy-variance-over-kF} we show the variance of the YSR energy for two dimensions. The numerical results confirm that the fluctuations become small in the limit of large $k_{\rm F} \ell$, while keeping $\ell/\xi_{\varepsilon}$ constant, quantitatively consistent with the result of lowest-order perturbation theory in the disorder potential $U(\vr)$.  Logarithmic corrections to the perturbative results are expected to occur deep in the dirty limit, see Eq.~(\ref{eq:2D-prediction-multipleScattering}).

\begin{figure}
\noindent\begin{minipage}[t]{1\columnwidth}%
\includegraphics[width=1\columnwidth]{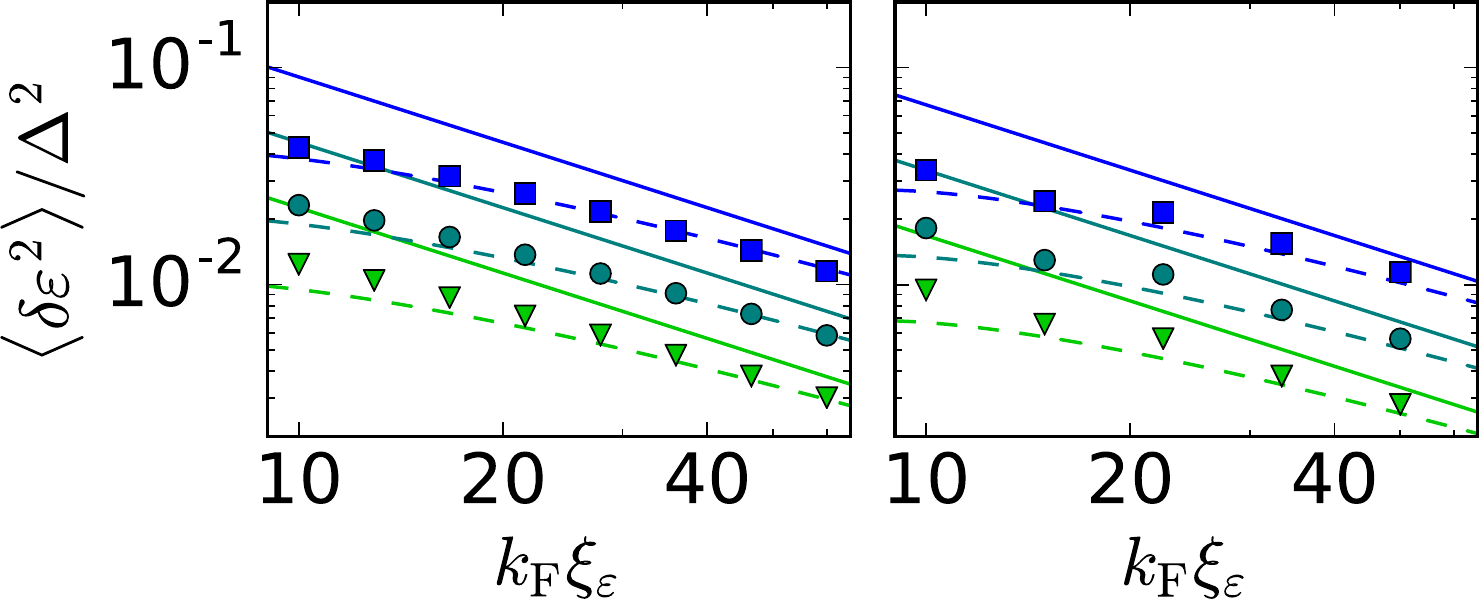}%
\end{minipage}\caption{YSR energy variance in three dimensions, as a function of $k_{\rm F} \xi_\varepsilon$ at fixed ratios $\xi_\varepsilon/\ell$. The parameters are the same as in Fig. \ref{fig:2D Energy-variance-over-kF}. The dashed and solid curves give the perturbative result (\ref{eq:3D-quant-pertTheory-higherOrders}) and its leading term, respectively.\label{fig:3D Energy-variance-over-kF}}
\end{figure}

For comparison with the numerical results in three dimensions, we have repeated the
perturbative calculation of Sec. \ref{sec:Perturbative-approach} with the disorder potential set to zero for $r < r_0 = 1/\kF$. In this case we find

\begin{align}\label{eq:3D-quant-pertTheory-higherOrders}
  &\langle \delta \varepsilon^2 \rangle = \frac{\Delta^2 (\xi_0/\xi_{\varepsilon_0})^4}{k_{\rm F}\ell} \\
  & \times \left(c_0  + \frac{c_1 - 2 \log(\kF \xie)}{\kF \xie} - \frac{c_3 + c_{4} \cos \eta}{\kF^2 \xie^2}  + \cdots \right)\nonumber
\end{align}
to second order in $1/\kF \xie$. The coefficients read $c_0 \approx 0.45$, $c_1 \approx 2.59$, $c_3 \approx 5.87$ and $c_{4} \approx 2.71$.  To leading order this simplifies
to the asymptotic form in Eq. (\ref{eq:ballistic - 3D integral}).
The agreement of the higher order result (\ref{eq:3D-quant-pertTheory-higherOrders}) with the numerics is excellent for all values of $k_{\rm F} \xi_{\varepsilon}$ considered;
the leading order agrees for large values of $k_{\rm F} \xi_{\varepsilon}$ only, see Fig. \ref{fig:3D Energy-variance-over-kF}. Additional data for parameters deeper in the dirty limit and at fixed $\kF$ and $\xi_0$ are provided in appendix~\ref{app:dirty_limit}.

\begin{figure}
\noindent\begin{minipage}[t]{1\columnwidth}%
\includegraphics[width=1\columnwidth]{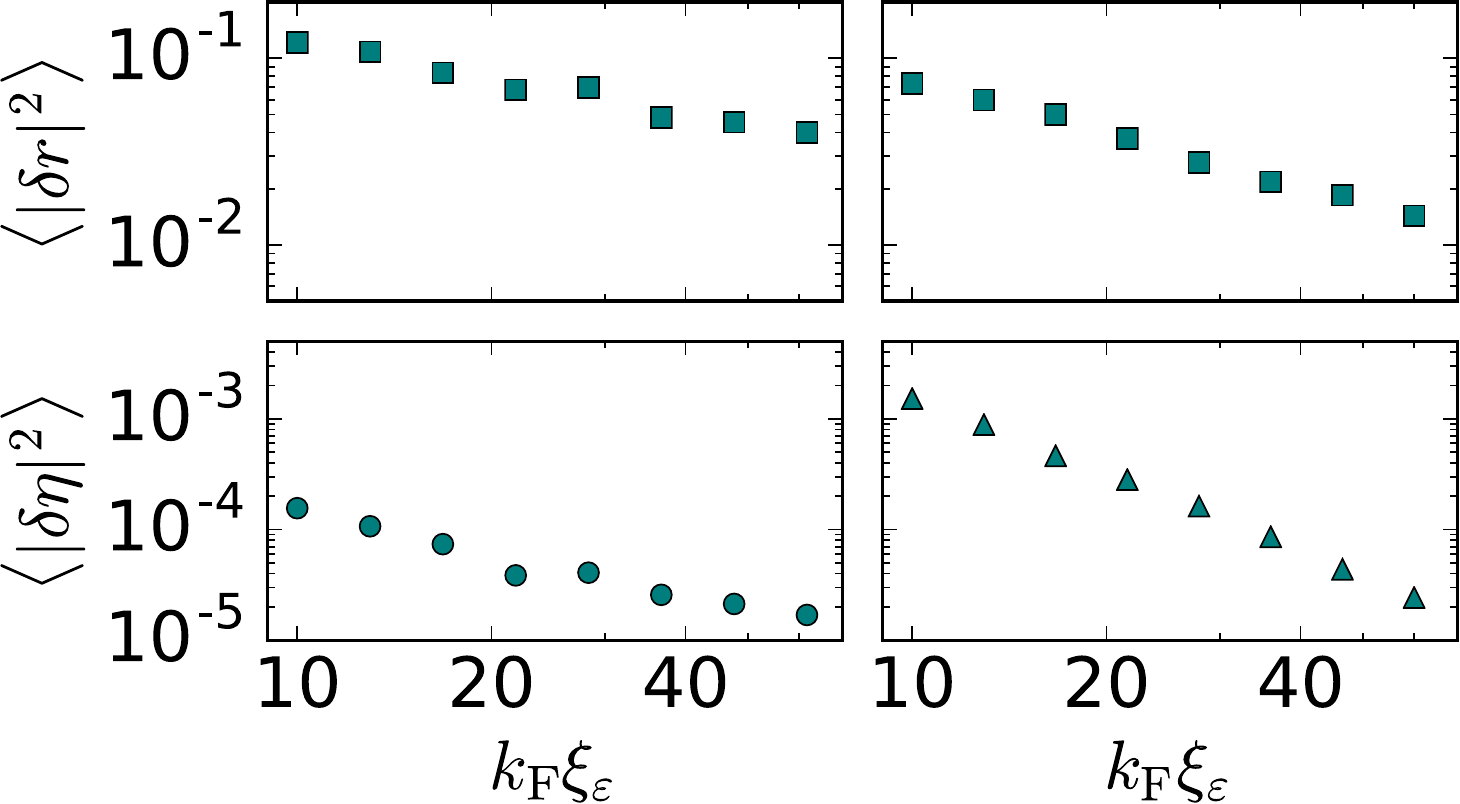}%
\end{minipage}\caption{\label{fig:Evar contributions}Comparison of the two contributions $\delta r$ and $\delta \eta$
to the superconductor scattering matrix $S(\varepsilon)$, in two (left) and three (right) dimensions. See Eq. (\ref{eq:dA definition}).
We choose a ratio $\xi_0/\ell=1$ and energy $\varepsilon/\Delta = 0.15$. The contribution from normal reflection dominates in two as well as in three dimensions.}
\end{figure}

Our approach also allows us to separate contributions to the YSR energy variance arising from fluctuations of the Andreev phase and normal reflection, see Eq.\ (\ref{eq:der}). The two contributions to $\langle \delta \varepsilon^2 \rangle$ are shown in Fig.\ \ref{fig:Evar contributions} for $\varepsilon/\Delta = 0.15$. The figure shows that the main contribution to $\langle \delta \varepsilon^2 \rangle$ comes from normal reflection. This is consistent with the fact that fluctuations of the Andreev phase $\eta(\varepsilon)$ have an additional smallness because $\delta \eta(\varepsilon)$ is an antisymmetric function of energy.

\section{Conclusion\label{sec:Conclusion}}

In this work we have analyzed the variance of the YSR energy due to
nonmagnetic disorder in both two and three dimensions. Mapping this
problem to a scattering ansatz for electrons and holes close to the
magnetic impurity allowed us to reduce the effects
of nonmagnetic disorder to two separate contributions.

First, the Andreev phase, which is picked up when an electron
is Andreev reflected as a hole or vise versa, starts fluctuating as a function of disorder configuration. Using time reversal
and particle hole symmetry, we have argued that this contribution
is expected to be negligible in the limit $k_{\text{F}} \ell ,\,k_{F}\xi_{\varepsilon}\gg1$. Numerical calculations confirmed this prediction.

Second, disorder can flip the momentum and lead to a finite normal reflection
amplitude for electrons or holes propagating from the impurity into the superconductor.
We find that the normal reflection probability is small if $k_{\rm F} \ell \gg \log \left[ k_{\rm F} \xi_\varepsilon \max(1,\xi_\epsilon/\ell)\right]$ and $k_{\rm F} \ell \gg 1$ in two and three dimensions, respectively. Importantly YSR states can be robust to disorder even in the limit of a dirty superconductor.

%Typical experimental parameters involve few $\si{\angstrom}$ and $\xi_0$ of order $\xi_0 \sim 40$ (Nb) to $\xi_0 \sim 1500$ (Al). The mean-free path $\ell$ depends on details such as the growth process. In order for the YSR energies to be robust to disorder, in three dimensions, $\ell$ simply has to be large compared to $\lambda_{\rm F}$, which is typically of order of a few $\si{\angstrom}$.  The logarithm in two dimensions sets higher requirements for large coherence length system similar to Aluminium.

Finally, we found that in three dimensions only disorder within
a few Fermi wavelengths of the magnetic impurity contributes to the
YSR energy variance. This is in contrast to two dimensions, where
disorder from distances up to the coherence length contributes.

Our results relax earlier claims \cite{Hui_2015_2}, which suggested that at $\ell$ of the order of $\xi_0$ the width of YSR energy distribution becomes of the order of the superconducting gap. Our findings show that $\lambda_{\rm F}$ is a crucial parameter to be included into the discussion and that this typically leads to a negligible influence of disorder.

Our findings also have implications for one dimensional topological superconductors, that are formed by dilute classical adatom chains. These systems can be described by effective tight-binding models \cite{Pientka2015}. (Note, however, that current experiments may well be in a rather different limit in which the hybridization of the adatom d levels plays an important role \cite{yazdani2014, franke2015, feldman2017,  meyer2016, franke2017}.) The on-site energies in these models are immune to disorder, if the conditions are met that we derived in this work for the single Shiba states. This leaves the discussion of the tunneling and pairing strengths. If the distance $d$ between the impurities is small compared to $\ell$, we expect the influence of disorder on the nearest neighbor hopping and pairing terms to be suppressed by a factor $d/\ell \sim 1/k_{\rm F}\ell$. However, previous studies have shown \cite{Pientka_2013, Pientka2014}, that in a clean system with $k_{\rm F} \xi_0 \gg k_{\rm F} d\gg 1$, tunneling is following a long-range, $1/r$ power law. Thus, strictly speaking, more than nearest-neighbor terms have to be included. If the length of the chain is smaller than the mean-free path, the same arguments as for the nearest-neighbor elements apply. For longer chains, further work is required to investigate the influence of disorder on long range tunneling and pairing elements.

\begin{acknowledgements}
We thank Pascal Simon, Yang Peng, and Bj\"orn Sbierski for discussions. Financial support was provided by the Institute ``Quantum Phenomenon in Novel Materials'' at the Helmholtz Zentrum Berlin, the Helmholtz Virtual Institute ``New States of Matter and Their Excitations'', and the Deutsche Forschungsgemeinschaft (CRC 183).
\end{acknowledgements}

\appendix
\renewcommand\thefigure{\thesection.\arabic{figure}} 
\renewcommand\appendixname{\uppercase{appendix}}
\newcommand{\appendixSection}[2]{\section{\uppercase{#1}#2}}

\appendixSection{Relation to Ref. }{\cite{Hui_2015_2}}\label{app:compHui}

In this section, we discuss the related Ref. \cite{Hui_2015_2}, which reports a much stronger susceptibility of the YSR energy to disorder than we do. We attribute the difference to the two approximations made in Ref. \cite{Hui_2015_2}. 

Without disorder, the magnetic impurity contributes a delta function $\delta(\omega - \varepsilon_0)$ to the density of states. Including and averaging over disorder, this contribution is broadened. In Ref. \cite{Hui_2015_2}, the width of the peak in the density of states is used as a measure of the disorder-induced variance $\langle \delta \varepsilon^2 \rangle$ of the YSR energy.

Reference \cite{Hui_2015_2} calculates the Green function $G(\omega)$ in the presence of the magnetic impurity and non-magnetic disorder in the superconductor and obtains the impurity density of states from the relation
\begin{equation}
\rho(\omega) = - \frac{1}{\pi} {\rm Tr}{\rm Im} \langle G(\omega) \rangle,
\end{equation}
where the brackets $\langle \cdots \rangle$ refer to the disorder average. The disorder average $\Gfull$ is then performed with two approximations. First, Ref. \cite{Hui_2015_2} uses the self-consistent Born approximation (SCBA), which yields a self-consistent equation for $\Gfull$,
\begin{equation}\label{eq:green_full}
	\Gfull = \left\{
			\left[\Go\right]^{-1} 
			- \Sigma(\omega)
		\right\}^{-1},
\end{equation}
where $\Go$ is the Green function without the non-magnetic disorder (but with the magnetic impurity) and
\begin{equation}\label{eq:single_sum}
	\Sigma_{\bf k, k'}(\omega) = 
		\frac{v_{\rm F}}{2\pi\nu_0\ell V} \sum_{\bf p} 
			\tau_z \left\langle G_{\bf p + k, p+k'}(\omega) \right \rangle \tau_z
\end{equation}
is the SCBA self energy (other symbols are defined in Sec. \ref{sec:Model}).

The second approximation in Ref. \cite{Hui_2015_2} is based on the following argument. The main contribution of the summation in the self energy is from momenta $\bf p + k$ and $\bf p + k'$ at the Fermi-level. Hence, approximately, one can restrict the summation over $\bf p$ to the manifold defined by ${|\bf p+k| = |p + k'|} = k_{\rm F}$. If $\vk \neq \vk'$, these are two independent equations and hence the manifold has two dimensions less than the dimension of ${\bf p}$. If however  $\vk =\vk'$, there is only one constraint and the dimension of the manifold is only one less than the dimension of ${\bf p}$. From this, the authors of Ref. \cite{Hui_2015_2} concluded that the self energy can be approximated to be diagonal and that it reads
\begin{equation}\label{eq:double_sum}
	\Sigma_{\vk,\vk'} \approx  \frac{v_{\rm F}}{2\pi\nu_0\ell V} \delta_{\vk,\vk'} 
		\sum_{\bf p, q} \tau_z \langle G_{\bf p, q} \rangle \tau_z.
\end{equation}

To facilitate the comparison with our own results, we first reformulate these in Green function language. In the limit $\kF \ell \gg \kF \xi_0 \gg 1$ the YSR state is separated from other states by a finite gap. Hence, only the lowest order contributions in disorder are expected to contribute to a shift in the YSR energy and to a good (controlled) approximation, we can rewrite the low-energy part of Hamiltonian \eqref{eq:general} as
\begin{equation}\label{eq:low_energy}
	H = (\varepsilon_0 + U_{0,0} )\ket{0}\bra{0}.
\end{equation}
Here, $\ket{0}$ is the YSR-state derived in the main text, with its wave-function given by Eq. (\ref{eq:general}). The disorder matrix-element $U_{0,0} =\bra{0} U(\vr) \tau_z \ket{0}$  has a Gaussian distribution with zero average and a variance $\langle U_{0,0}^2 \rangle =\langle \delta \varepsilon^2\rangle$, where the latter was derived in Eqs. (\ref{eq:2D - 1st order result}) and (\ref{eq:ballistic - 3D integral}) in the main text. Note that, due to particle-hole symmetry being present in the physical problem, there is also a YSR state at $-\varepsilon_0$. However, this second state lives in a disjunct sector of the Hilbert space and hence it is sufficient to consider only one of the two states when calculating the spectrum.

The Green function is easily obtained and, within the YSR-state subspace, reads
\begin{equation}\label{eq:green_gaussian}
        G(\omega) = \frac{1}{\omega - \varepsilon_0 - U_{0,0} + i \eta}
\end{equation}
with $\eta \downarrow 0$. The average density of states reads
\begin{equation}\label{eq:dos_gaussian}
        \rho(\omega) = \frac{1}{\sqrt{2\pi \langle\delta \varepsilon\rangle^2}}e^{-(\omega - \varepsilon_0)^2/2 \langle\delta \varepsilon\rangle^2},
\end{equation}
in exact quantitative agreement with the perturbation theory of Sec. \ref{sec:Perturbative-approach}.

We now investigate the effect of the first approximation in \cite{Hui_2015_2}. For the low-energy Hamiltonian \eqref{eq:low_energy} the expression for the SCBA self energy reads
\begin{equation}\label{eq:self_energy_SCBA}
        \Sigma(\omega) = \langle \delta \varepsilon^2 \rangle \Gfull.
\end{equation}
Solving Eqs. \eqref{eq:green_full} and \eqref{eq:self_energy_SCBA} one finds 
\begin{align}
	\rho(\omega) = \begin{cases}
		\frac{\sqrt{4 \langle \delta \varepsilon^2 \rangle - (\omega - \varepsilon_0)^2}}{2\pi \langle \delta \varepsilon^2\rangle}
		& \text{for } (\omega - \varepsilon_0)^2 < 4 \langle \delta \varepsilon^2 \rangle, \\
		0 & \text{else.}
	\end{cases}
\end{align}
This result disagrees qualitatively from the exact result \eqref{eq:dos_gaussian}, although the order-of-magnitude of the width of the density of states peak is still correct.

The second approximation relies on the assumption that $\Sigma(\omega)$ and $\Gfull$ are diagonal in momentum space. The latter assumption is clearly questionable, since the presence of the magnetic impurity causes the Green function to be non-diagonal in momentum space. In contrast,  replacing the single sum in Eq. \eqref{eq:single_sum} by the double sum in Eq. \eqref{eq:double_sum} assumes a diagonal Green function. Taking the double sum greatly increases the impurity contribution to the diagonal part of $\Sigma(\omega)$, whereas the approximation \eqref{eq:double_sum} ignores any impurity-induced off-diagonal contribution to $\Sigma(\omega)$. With this second approximation, the momentum sums can be replaced by energy integrations. In this step, the dependence on $\kF \ell$ as well as system dimensionality drops out, leaving a dependence on the ratio $\ell/\xi_0$ only, a feature that clearly contradicts the direct perturbative solution \eqref{eq:dos_gaussian} in the weak-disorder limit $\xi_0 \ll \ell$.

To conclude, while both approximation made in Ref. \cite{Hui_2015_2} are uncontrolled, we believe it is the second approximation that is responsible for the stark qualitative difference between that reference and the present results.

\appendixSection{Energy variance as a function of mean-free path}{}\label{app:dirty_limit}
\setcounter{figure}{0}   

\begin{figure}
\includegraphics[width=1\columnwidth]{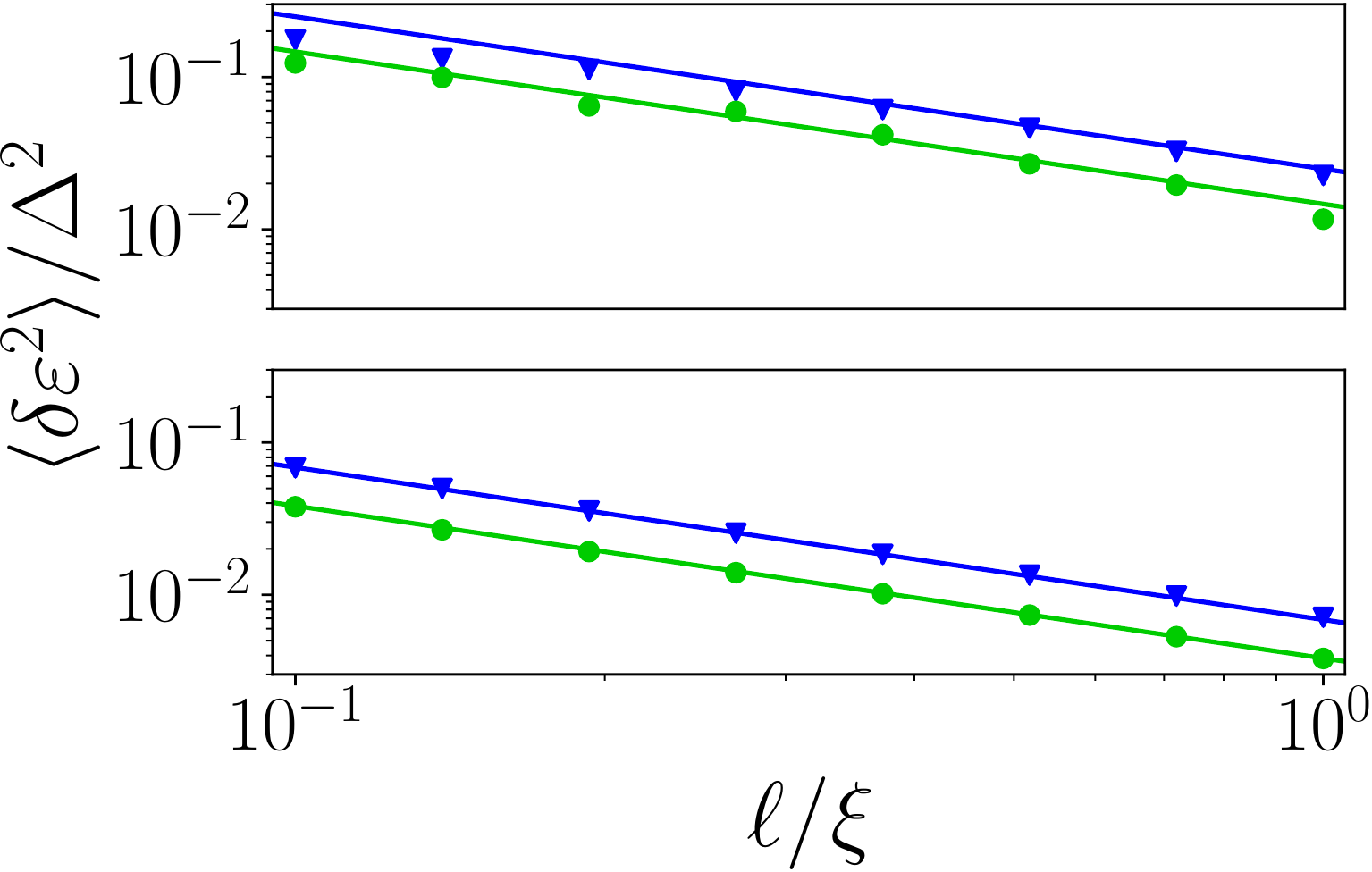}
\caption{\label{fig:mean_free_path_sweep}
Energy variance over mean-free path at fixed $\kF$ and $\xi_0$. The markers show the numerical results in two dimensions (top) and three dimensions (bottom), with $\kF \xi_0 = 50$ (blue triangles) and $\kF \xi_0 = 100$ (green circles). The lines correspond to the perturbative results derived in the main text.  The scattering phase shifts $\phi_\uparrow = - \phi_\downarrow$ of the magnetic impurity are chosen such that, in the absence of disorder, a YSR state forms at $\varepsilon / \Delta = 0$.
} 
\end{figure}

In this section we supplement the numerical results of the main text by data in the dirty limit, $\ell \lesssim \xi_0$, and for a fixed $\xi_0$ and $\kF$ while varying $\ell$. The data is shown in Fig. \ref{fig:mean_free_path_sweep}, with the perturbative results taken from Eqs. (\ref{eq:2D - 1st order result}) and (\ref{eq:3D-quant-pertTheory-higherOrders}). The energy variance is well approximated by the lowest order perturbation theory in disorder. Deviations occur in two dimensions, when $\kF \ell$ gets close to one.

\begin{figure}
\includegraphics[width=1\columnwidth]{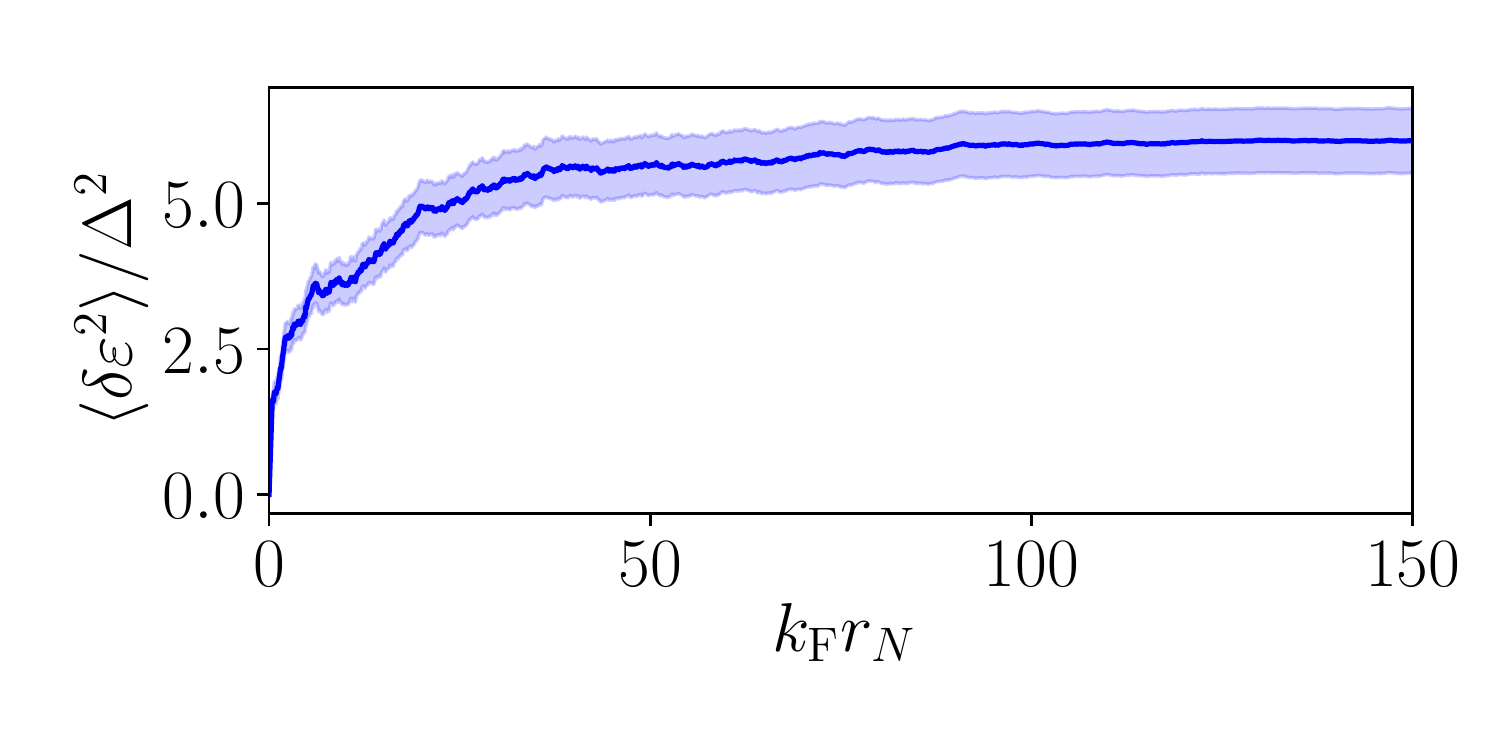}
\caption{\label{fig:dirty_limit_conv}
Convergence of the energy variance with disorder cutoff $r_N$ in the dirty limit and in two dimensions. The parameters are $\kF \xi_0 = 300$ and $\xi_0/\ell = 10$, with the same magnetic impurity parameters as in Fig. \ref{fig:mean_free_path_sweep}.
} 
\end{figure}

Additionally, as shown in Fig. \ref{fig:convergence}, in order for us to reach convergence in two dimensions we had to let the disorder cutoff flow to a distance far exceeding the Fermi wave length. Here, we supplement the main text plot by a fully converged plot in the dirty limit, shown in Fig. \ref{fig:dirty_limit_conv}. We note especially, that convergence requires distances of the order of multiple mean-free paths and thus the final value is expected to contain contributions from multiple scattering.

\appendixSection{Transfer matrix for a single slice}{}
\setcounter{figure}{0}   

In this section we present explicit expressions for the transfer matrix of a thin, disordered and superconducting slice. Using Eq. (\ref{eq:numerical Sn}) this enables one to calculate the corresponding scattering matrix of the thin slice.

\subsection{Two dimensions}

First we consider a two dimensional, circular slice. In this case, we define the radial part of the propagating waves in Eq. (\ref{eq:2D - modes}) as
\[
\ket{{\cal R}_{\text{e}m\pm}(r)}=\sqrt{\frac{k_{\rm F}}{4 v_{\rm F}}}H_{m}^{(\pm)}(k_{\rm F}r)\begin{pmatrix}1\\
0
\end{pmatrix},
\]
for electrons and 
\[
\ket{{\cal R}_{\text{h}m\pm}(r)}=\sqrt{\frac{k_{\rm F}}{4 v_{\rm F}}}H_{m}^{(\mp)}(k_{\rm F}r) \begin{pmatrix}0\\
1
\end{pmatrix},
\]
for holes.

Within first order Born approximation, for a slice of width $\rm{d}r$, at an energy $\varepsilon$
and at a radius $r_n$ from the origin, we obtain the transfer matrix
\[
{\cal T}^{(n)}_{\nu'\nu}= \bra{{\cal R}_{\nu'}(r_n)}
\\
\mathrm{d} \hat{T}^{(n)} \delta_{m',m}
+
\mathrm{d}\hat{\Gamma}_{m',\,m}^{(n)}
\ket{{\cal R}_\nu(r_n)},\label{eq:transfer matrix 2D}
\]
with the same multi-index $\nu$ as in the main text. The term diagonal in angular momentum reads
\[
\mathrm{d} \hat{T}^{(n)}=(\Delta \tau_x - \epsilon) 2 \pi r_n {\rm d} r .
\]
For the disorder element we get
\[
\mathrm{d}\hat{\Gamma}_{m',\,m}^{(n)} = 
\sqrt{\gamma2\pi r_n\mathrm{d}r}X_{m'-m}^{(n)}\tau_{z}.
\]
The random part is absorbed into 
\[
X_{m'-m}^{(n)}=\begin{cases}
Y_{0}^{(n)} & \text{for }\text{m=m'}\\
\frac{Y_{|m'-m|}^{(n)}+i\mathrm{sign}(m'-m)Z_{|m'-m|}^{(n)}}{\sqrt{2}} & \text{else}
\end{cases}
\]
corresponding to the Fourier transform of white noise. Here $Y_{m}^{(n)}$
and $Z_{m}^{(n)}$ are independent, normally distributed random
variables with zero mean and variance one. Note that not all elements
of $X^{(n)}$ are independent since $X_{m'-m}=X_{m-m'}^{*}$.

\subsection{Three dimensions}

Next we calculate the transfer matrix of a thin, spherical slice in three dimensions. In this case, the radial modes are defined as
\[
\ket{{\cal R}_{\text{e}m\pm}(r)} = \sqrt{\frac{k_{\rm F}^2}{4\pi v_{\rm F}}} h_l^{(\pm)}(k_{\rm F} r) \begin{pmatrix}
1\\0
\end{pmatrix}
\]
for electrons and

\[
\ket{{\cal R}_{\text{h}m\pm}(r)} = \sqrt{\frac{k_{\rm F}^2}{4\pi v_{\rm F}}} h_l^{(\mp)}(k_{\rm F} r) \begin{pmatrix}
0\\1
\end{pmatrix}
\]
for holes, with a total angular momentum quantum number $l$.

Within first order Born approximation and for a spherical slice of width ${\rm d}r$ at a radius $r_n$, the transfer matrix reads 
\[
{\cal T}^{(n)}_{\nu'\nu}= \bra{{\cal R}_{\nu'}(r_n)}
\\
\mathrm{d} \hat{T}^{(n)} \delta_{l'l}\delta_{m'm}
+
\mathrm{d}\hat{\Gamma}_{l'm'lm}^{(n)}
\ket{{\cal R}_\nu(r_n)}
\]
similar to Eq. (\ref{eq:transfer matrix 2D}).

The term diagonal in the angular momentum quantum numbers reads
\[
{\rm d}\hat{T}^{(n)} = (\Delta \tau_x - \varepsilon) 4 \pi r_n^2 {\rm d}r
\]
and the disorder element is
\[
\mathrm{d}\hat{\Gamma}_{l'm'lm}^{(n)} = \tau_z \sqrt{\gamma 4 \pi r_n^2 {\rm d} r} \Xi_{l'm'lm}.
\]

The random variable $\Xi_{l'm'lm}$ takes a more complicated form in three than in two dimensions, due to the involvement of the product of two spherical harmonics in the calculation of the matrix elements in Eq. (\ref{eq:Tmatrix definition}). These products can be expressed as a sum over single spherical harmonics for which explicit expressions are known \cite{varshalovich1988}. For these single spherical harmonics we can calculate the overlap with the Gaussian disorder potential. Following this strategy we obtain
\[
\Xi_{l'm'lm} = \sum_{L=0}^{\infty} c^{L,\,\Delta m}_{l'm'lm} X_{L,\,\Delta m}
\]
where the coefficients for the transformation between a single and the product of two spherical harmonics are
\begin{align}
c^{L,\,\Delta m}_{l'm'lm} =& (-1)^m \sqrt{\frac{ (2l'+1)(2l+1)}{4\pi (2 L + 1)}} \\
&\times C_{l',\,0,\,l,\,0}^{L,\,0} C_{l',\,m',\,l,\,m}^{L,\,\Delta m}.
\end{align}
Here 
\[
C_{l',\,m',\,l,\,m}^{L,\,\Delta m} = \braket{l',\,m',\,l,\,m|L,\,\Delta m}
\]
are the Clebsch Gordan coefficients \cite{varshalovich1988}.
Additionally, the overlap of the single spherical harmonics with the angular part of the Gaussian disorder potential yields the random numbers \cite{lang2015}
\[
X_{L,\,\Delta m}^{(n)}=\begin{cases}
\frac{Y_{L,\,\Delta m}^{(n)}+i Z_{L,\,\Delta m}^{(n)}}{\sqrt{2}} & \text{for } \Delta m >0\\

Y_{L,\,0}^{(n)} & \text{for }\Delta m = 0\\
(-1)^{\Delta m} X_{L,\,-\Delta m}^{(n)} & \text{for } \Delta m < 0
\end{cases}
\]
Here, similar to the two dimensional case, $Y_{L,\,\Delta m}$ and $Z_{L,\,\Delta m}$ are independent, normally distributed random variables with zero mean and variance one.

\FloatBarrier\bibliographystyle{apsrev4-1}
\bibliography{references}

%merlin.mbs apsrev4-1.bst 2010-07-25 4.21a (PWD, AO, DPC) hacked
%Control: key (0)
%Control: author (72) initials jnrlst
%Control: editor formatted (1) identically to author
%Control: production of article title (-1) disabled
%Control: page (0) single
%Control: year (1) truncated
%Control: production of eprint (0) enabled
\begin{thebibliography}{46}%
\makeatletter
\providecommand \@ifxundefined [1]{%
 \@ifx{#1\undefined}
}%
\providecommand \@ifnum [1]{%
 \ifnum #1\expandafter \@firstoftwo
 \else \expandafter \@secondoftwo
 \fi
}%
\providecommand \@ifx [1]{%
 \ifx #1\expandafter \@firstoftwo
 \else \expandafter \@secondoftwo
 \fi
}%
\providecommand \natexlab [1]{#1}%
\providecommand \enquote  [1]{``#1''}%
\providecommand \bibnamefont  [1]{#1}%
\providecommand \bibfnamefont [1]{#1}%
\providecommand \citenamefont [1]{#1}%
\providecommand \href@noop [0]{\@secondoftwo}%
\providecommand \href [0]{\begingroup \@sanitize@url \@href}%
\providecommand \@href[1]{\@@startlink{#1}\@@href}%
\providecommand \@@href[1]{\endgroup#1\@@endlink}%
\providecommand \@sanitize@url [0]{\catcode `\\12\catcode `\$12\catcode
  `\&12\catcode `\#12\catcode `\^12\catcode `\_12\catcode `\%12\relax}%
\providecommand \@@startlink[1]{}%
\providecommand \@@endlink[0]{}%
\providecommand \url  [0]{\begingroup\@sanitize@url \@url }%
\providecommand \@url [1]{\endgroup\@href {#1}{\urlprefix }}%
\providecommand \urlprefix  [0]{URL }%
\providecommand \Eprint [0]{\href }%
\providecommand \doibase [0]{http://dx.doi.org/}%
\providecommand \selectlanguage [0]{\@gobble}%
\providecommand \bibinfo  [0]{\@secondoftwo}%
\providecommand \bibfield  [0]{\@secondoftwo}%
\providecommand \translation [1]{[#1]}%
\providecommand \BibitemOpen [0]{}%
\providecommand \bibitemStop [0]{}%
\providecommand \bibitemNoStop [0]{.\EOS\space}%
\providecommand \EOS [0]{\spacefactor3000\relax}%
\providecommand \BibitemShut  [1]{\csname bibitem#1\endcsname}%
\let\auto@bib@innerbib\@empty
%</preamble>
\bibitem [{\citenamefont {Yu}(1965)}]{YU1965}%
  \BibitemOpen
  \bibfield  {author} {\bibinfo {author} {\bibfnamefont {L.}~\bibnamefont
  {Yu}},\ }\href {\doibase 10.7498/aps.21.75} {\bibfield  {journal} {\bibinfo
  {journal} {Acta Phys. Sin.}\ }\textbf {\bibinfo {volume} {21}},\ \bibinfo
  {eid} {75} (\bibinfo {year} {1965})}\BibitemShut {NoStop}%
\bibitem [{\citenamefont {Shiba}(1968)}]{Shiba_1968}%
  \BibitemOpen
  \bibfield  {author} {\bibinfo {author} {\bibfnamefont {H.}~\bibnamefont
  {Shiba}},\ }\href {\doibase 10.1143/ptp.40.435} {\bibfield  {journal}
  {\bibinfo  {journal} {Prog. Theor. Phys.}\ }\textbf {\bibinfo {volume}
  {40}},\ \bibinfo {pages} {435} (\bibinfo {year} {1968})}\BibitemShut
  {NoStop}%
\bibitem [{\citenamefont {Rusinov}(1968)}]{1969JETPL...9...85R}%
  \BibitemOpen
  \bibfield  {author} {\bibinfo {author} {\bibfnamefont {A.~I.}\ \bibnamefont
  {Rusinov}},\ }\href@noop {} {\bibfield  {journal} {\bibinfo  {journal} {Zh.
  Eksp. Teor. Fiz. Pisma Red.}\ }\textbf {\bibinfo {volume} {9}},\ \bibinfo
  {pages} {146} (\bibinfo {year} {1968})},\ \bibinfo {note} {[JETP Lett.\ {\bf
  9}, 85 (1969)]}\BibitemShut {NoStop}%
\bibitem [{\citenamefont {Rusinov}(1969)}]{1969JETP...29.1101R}%
  \BibitemOpen
  \bibfield  {author} {\bibinfo {author} {\bibfnamefont {A.~I.}\ \bibnamefont
  {Rusinov}},\ }\href@noop {} {\bibfield  {journal} {\bibinfo  {journal} {Zh.
  Eksp. Teor. Fiz. Pisma Red.}\ }\textbf {\bibinfo {volume} {56}},\ \bibinfo
  {pages} {2047} (\bibinfo {year} {1969})},\ \bibinfo {note} {[JETP Lett.\ {\bf
  29}, 1101 (1969)]}\BibitemShut {NoStop}%
\bibitem [{\citenamefont {Zittartz}\ \emph {et~al.}(1972)\citenamefont
  {Zittartz}, \citenamefont {Bringer},\ and\ \citenamefont
  {M{\"u}ller-Hartmann}}]{Zittartz_1972}%
  \BibitemOpen
  \bibfield  {author} {\bibinfo {author} {\bibfnamefont {J.}~\bibnamefont
  {Zittartz}}, \bibinfo {author} {\bibfnamefont {A.}~\bibnamefont {Bringer}}, \
  and\ \bibinfo {author} {\bibfnamefont {E.}~\bibnamefont
  {M{\"u}ller-Hartmann}},\ }\href {\doibase 10.1016/0038-1098(72)90056-7}
  {\bibfield  {journal} {\bibinfo  {journal} {Solid State Commun.}\ }\textbf
  {\bibinfo {volume} {10}},\ \bibinfo {pages} {513 } (\bibinfo {year}
  {1972})}\BibitemShut {NoStop}%
\bibitem [{\citenamefont {Hirschfeld}\ \emph {et~al.}(1988)\citenamefont
  {Hirschfeld}, \citenamefont {W\"olfle},\ and\ \citenamefont
  {Einzel}}]{Shiba_Cite_Hirschfeld_1988}%
  \BibitemOpen
  \bibfield  {author} {\bibinfo {author} {\bibfnamefont {P.~J.}\ \bibnamefont
  {Hirschfeld}}, \bibinfo {author} {\bibfnamefont {P.}~\bibnamefont
  {W\"olfle}}, \ and\ \bibinfo {author} {\bibfnamefont {D.}~\bibnamefont
  {Einzel}},\ }\href {\doibase 10.1103/physrevb.37.83} {\bibfield  {journal}
  {\bibinfo  {journal} {Phys. Rev. B}\ }\textbf {\bibinfo {volume} {37}},\
  \bibinfo {pages} {83} (\bibinfo {year} {1988})}\BibitemShut {NoStop}%
\bibitem [{\citenamefont {Balatsky}\ \emph {et~al.}(2006)\citenamefont
  {Balatsky}, \citenamefont {Vekhter},\ and\ \citenamefont
  {Zhu}}]{Shiba_Cite_Balatsky_2006}%
  \BibitemOpen
  \bibfield  {author} {\bibinfo {author} {\bibfnamefont {A.~V.}\ \bibnamefont
  {Balatsky}}, \bibinfo {author} {\bibfnamefont {I.}~\bibnamefont {Vekhter}}, \
  and\ \bibinfo {author} {\bibfnamefont {J.-X.}\ \bibnamefont {Zhu}},\ }\href
  {https://journals.aps.org/rmp/abstract/10.1103/RevModPhys.78.373} {\bibfield
  {journal} {\bibinfo  {journal} {Rev. Mod. Phys.}\ }\textbf {\bibinfo {volume}
  {78}},\ \bibinfo {pages} {373} (\bibinfo {year} {2006})}\BibitemShut
  {NoStop}%
\bibitem [{\citenamefont {Morr}\ and\ \citenamefont {Y.}(2006)}]{Morr2006}%
  \BibitemOpen
  \bibfield  {author} {\bibinfo {author} {\bibfnamefont {D.~K.}\ \bibnamefont
  {Morr}}\ and\ \bibinfo {author} {\bibfnamefont {J.}~\bibnamefont {Y.}},\
  }\href {\doibase 10.1103/PhysRevB.73.224511} {\bibfield  {journal} {\bibinfo
  {journal} {Phys. Rev. B}\ }\textbf {\bibinfo {volume} {73}},\ \bibinfo
  {pages} {224511} (\bibinfo {year} {2006})}\BibitemShut {NoStop}%
\bibitem [{\citenamefont {Moca}\ \emph {et~al.}(2008)\citenamefont {Moca},
  \citenamefont {Demler}, \citenamefont {Jank\'o},\ and\ \citenamefont
  {Zar\'and}}]{PhysRevB.77.174516}%
  \BibitemOpen
  \bibfield  {author} {\bibinfo {author} {\bibfnamefont {C.~P.}\ \bibnamefont
  {Moca}}, \bibinfo {author} {\bibfnamefont {E.}~\bibnamefont {Demler}},
  \bibinfo {author} {\bibfnamefont {B.}~\bibnamefont {Jank\'o}}, \ and\
  \bibinfo {author} {\bibfnamefont {G.}~\bibnamefont {Zar\'and}},\ }\href
  {\doibase 10.1103/PhysRevB.77.174516} {\bibfield  {journal} {\bibinfo
  {journal} {Phys. Rev. B}\ }\textbf {\bibinfo {volume} {77}},\ \bibinfo
  {pages} {174516} (\bibinfo {year} {2008})}\BibitemShut {NoStop}%
\bibitem [{\citenamefont {\ifmmode~\check{Z}\else \v{Z}\fi{}itko}\ \emph
  {et~al.}(2011)\citenamefont {\ifmmode~\check{Z}\else \v{Z}\fi{}itko},
  \citenamefont {Bodensiek},\ and\ \citenamefont {Pruschke}}]{Zitko2011}%
  \BibitemOpen
  \bibfield  {author} {\bibinfo {author} {\bibfnamefont {R.}~\bibnamefont
  {\ifmmode~\check{Z}\else \v{Z}\fi{}itko}}, \bibinfo {author} {\bibfnamefont
  {O.}~\bibnamefont {Bodensiek}}, \ and\ \bibinfo {author} {\bibfnamefont
  {T.}~\bibnamefont {Pruschke}},\ }\href {\doibase 10.1103/PhysRevB.83.054512}
  {\bibfield  {journal} {\bibinfo  {journal} {Phys. Rev. B}\ }\textbf {\bibinfo
  {volume} {83}},\ \bibinfo {pages} {054512} (\bibinfo {year}
  {2011})}\BibitemShut {NoStop}%
\bibitem [{\citenamefont {Yao}\ \emph {et~al.}(2014)\citenamefont {Yao},
  \citenamefont {Glazman}, \citenamefont {Demler}, \citenamefont {Lukin},\ and\
  \citenamefont {Sau}}]{Yao2014}%
  \BibitemOpen
  \bibfield  {author} {\bibinfo {author} {\bibfnamefont {N.}~\bibnamefont
  {Yao}}, \bibinfo {author} {\bibfnamefont {L.}~\bibnamefont {Glazman}},
  \bibinfo {author} {\bibfnamefont {E.}~\bibnamefont {Demler}}, \bibinfo
  {author} {\bibfnamefont {M.}~\bibnamefont {Lukin}}, \ and\ \bibinfo {author}
  {\bibfnamefont {J.}~\bibnamefont {Sau}},\ }\href {\doibase
  10.1103/PhysRevLett.113.087202} {\bibfield  {journal} {\bibinfo  {journal}
  {Phys. Rev. Lett.}\ }\textbf {\bibinfo {volume} {113}},\ \bibinfo {pages}
  {087202} (\bibinfo {year} {2014})}\BibitemShut {NoStop}%
\bibitem [{\citenamefont {Hoffman}\ \emph {et~al.}(2015)\citenamefont
  {Hoffman}, \citenamefont {Klinovaja}, \citenamefont {Meng},\ and\
  \citenamefont {Loss}}]{Hoffman2015}%
  \BibitemOpen
  \bibfield  {author} {\bibinfo {author} {\bibfnamefont {S.}~\bibnamefont
  {Hoffman}}, \bibinfo {author} {\bibfnamefont {J.}~\bibnamefont {Klinovaja}},
  \bibinfo {author} {\bibfnamefont {T.}~\bibnamefont {Meng}}, \ and\ \bibinfo
  {author} {\bibfnamefont {D.}~\bibnamefont {Loss}},\ }\href {\doibase
  10.1103/PhysRevB.92.125422} {\bibfield  {journal} {\bibinfo  {journal} {Phys.
  Rev. B}\ }\textbf {\bibinfo {volume} {92}},\ \bibinfo {pages} {125422}
  (\bibinfo {year} {2015})}\BibitemShut {NoStop}%
\bibitem [{\citenamefont {Kim}\ \emph {et~al.}(2015)\citenamefont {Kim},
  \citenamefont {Zhang}, \citenamefont {Rossi},\ and\ \citenamefont
  {Lutchyn}}]{Kim2015}%
  \BibitemOpen
  \bibfield  {author} {\bibinfo {author} {\bibfnamefont {Y.}~\bibnamefont
  {Kim}}, \bibinfo {author} {\bibfnamefont {J.}~\bibnamefont {Zhang}}, \bibinfo
  {author} {\bibfnamefont {E.}~\bibnamefont {Rossi}}, \ and\ \bibinfo {author}
  {\bibfnamefont {R.~M.}\ \bibnamefont {Lutchyn}},\ }\href {\doibase
  10.1103/PhysRevLett.114.236804} {\bibfield  {journal} {\bibinfo  {journal}
  {Phys. Rev. Lett.}\ }\textbf {\bibinfo {volume} {114}},\ \bibinfo {pages}
  {236804} (\bibinfo {year} {2015})}\BibitemShut {NoStop}%
\bibitem [{\citenamefont {Meng}\ \emph {et~al.}(2015)\citenamefont {Meng},
  \citenamefont {Klinovaja}, \citenamefont {Hoffman}, \citenamefont {Simon},\
  and\ \citenamefont {Loss}}]{Meng2015}%
  \BibitemOpen
  \bibfield  {author} {\bibinfo {author} {\bibfnamefont {T.}~\bibnamefont
  {Meng}}, \bibinfo {author} {\bibfnamefont {J.}~\bibnamefont {Klinovaja}},
  \bibinfo {author} {\bibfnamefont {S.}~\bibnamefont {Hoffman}}, \bibinfo
  {author} {\bibfnamefont {P.}~\bibnamefont {Simon}}, \ and\ \bibinfo {author}
  {\bibfnamefont {D.}~\bibnamefont {Loss}},\ }\href {\doibase
  10.1103/PhysRevB.92.064503} {\bibfield  {journal} {\bibinfo  {journal} {Phys.
  Rev. B}\ }\textbf {\bibinfo {volume} {92}},\ \bibinfo {pages} {064503}
  (\bibinfo {year} {2015})}\BibitemShut {NoStop}%
\bibitem [{\citenamefont {\ifmmode~\check{Z}\else
  \v{Z}\fi{}itko}(2016)}]{Zitko2016}%
  \BibitemOpen
  \bibfield  {author} {\bibinfo {author} {\bibfnamefont {R.}~\bibnamefont
  {\ifmmode~\check{Z}\else \v{Z}\fi{}itko}},\ }\href {\doibase
  10.1103/PhysRevB.93.195125} {\bibfield  {journal} {\bibinfo  {journal} {Phys.
  Rev. B}\ }\textbf {\bibinfo {volume} {93}},\ \bibinfo {pages} {195125}
  (\bibinfo {year} {2016})}\BibitemShut {NoStop}%
\bibitem [{\citenamefont {Yazdani}\ \emph {et~al.}(1997)\citenamefont
  {Yazdani}, \citenamefont {Jones}, \citenamefont {Lutz}, \citenamefont
  {Crommie},\ and\ \citenamefont {Eigler}}]{Yazdani_1997}%
  \BibitemOpen
  \bibfield  {author} {\bibinfo {author} {\bibfnamefont {A.}~\bibnamefont
  {Yazdani}}, \bibinfo {author} {\bibfnamefont {B.~A.}\ \bibnamefont {Jones}},
  \bibinfo {author} {\bibfnamefont {C.~P.}\ \bibnamefont {Lutz}}, \bibinfo
  {author} {\bibfnamefont {M.~F.}\ \bibnamefont {Crommie}}, \ and\ \bibinfo
  {author} {\bibfnamefont {D.~M.}\ \bibnamefont {Eigler}},\ }\href {\doibase
  10.1126/science.275.5307.1767} {\bibfield  {journal} {\bibinfo  {journal}
  {Science}\ }\textbf {\bibinfo {volume} {275}},\ \bibinfo {pages} {1767}
  (\bibinfo {year} {1997})}\BibitemShut {NoStop}%
\bibitem [{\citenamefont {Ji}\ \emph {et~al.}(2008)\citenamefont {Ji},
  \citenamefont {Zhang}, \citenamefont {Fu}, \citenamefont {Chen},
  \citenamefont {Ma}, \citenamefont {Li}, \citenamefont {Duan}, \citenamefont
  {Jia},\ and\ \citenamefont {Xue}}]{Ji_2008}%
  \BibitemOpen
  \bibfield  {author} {\bibinfo {author} {\bibfnamefont {S.-H.}\ \bibnamefont
  {Ji}}, \bibinfo {author} {\bibfnamefont {T.}~\bibnamefont {Zhang}}, \bibinfo
  {author} {\bibfnamefont {Y.-S.}\ \bibnamefont {Fu}}, \bibinfo {author}
  {\bibfnamefont {X.}~\bibnamefont {Chen}}, \bibinfo {author} {\bibfnamefont
  {X.-C.}\ \bibnamefont {Ma}}, \bibinfo {author} {\bibfnamefont
  {J.}~\bibnamefont {Li}}, \bibinfo {author} {\bibfnamefont {W.-H.}\
  \bibnamefont {Duan}}, \bibinfo {author} {\bibfnamefont {J.-F.}\ \bibnamefont
  {Jia}}, \ and\ \bibinfo {author} {\bibfnamefont {Q.-K.}\ \bibnamefont
  {Xue}},\ }\href@noop {} {\bibfield  {journal} {\bibinfo  {journal} {Phys.
  Rev. Let.}\ }\textbf {\bibinfo {volume} {100}},\ \bibinfo {pages} {226801}
  (\bibinfo {year} {2008})}\BibitemShut {NoStop}%
\bibitem [{\citenamefont {Ji}\ \emph {et~al.}(2010)\citenamefont {Ji},
  \citenamefont {Zhang}, \citenamefont {Fu}, \citenamefont {Chen},
  \citenamefont {Jia}, \citenamefont {Xue},\ and\ \citenamefont
  {Ma}}]{Ji_2010}%
  \BibitemOpen
  \bibfield  {author} {\bibinfo {author} {\bibfnamefont {S.-H.}\ \bibnamefont
  {Ji}}, \bibinfo {author} {\bibfnamefont {T.}~\bibnamefont {Zhang}}, \bibinfo
  {author} {\bibfnamefont {Y.-S.}\ \bibnamefont {Fu}}, \bibinfo {author}
  {\bibfnamefont {X.}~\bibnamefont {Chen}}, \bibinfo {author} {\bibfnamefont
  {J.-F.}\ \bibnamefont {Jia}}, \bibinfo {author} {\bibfnamefont {Q.-K.}\
  \bibnamefont {Xue}}, \ and\ \bibinfo {author} {\bibfnamefont {X.-C.}\
  \bibnamefont {Ma}},\ }\href {\doibase 10.1063/1.3318404} {\bibfield
  {journal} {\bibinfo  {journal} {Appl. Phys. Let.}\ }\textbf {\bibinfo
  {volume} {96}},\ \bibinfo {pages} {073113} (\bibinfo {year}
  {2010})}\BibitemShut {NoStop}%
\bibitem [{\citenamefont {Franke}\ \emph {et~al.}(2011)\citenamefont {Franke},
  \citenamefont {Schulze},\ and\ \citenamefont {Pascual}}]{Franke940}%
  \BibitemOpen
  \bibfield  {author} {\bibinfo {author} {\bibfnamefont {K.~J.}\ \bibnamefont
  {Franke}}, \bibinfo {author} {\bibfnamefont {G.}~\bibnamefont {Schulze}}, \
  and\ \bibinfo {author} {\bibfnamefont {J.~I.}\ \bibnamefont {Pascual}},\
  }\href {\doibase 10.1126/science.1202204} {\bibfield  {journal} {\bibinfo
  {journal} {Science}\ }\textbf {\bibinfo {volume} {332}},\ \bibinfo {pages}
  {940} (\bibinfo {year} {2011})}\BibitemShut {NoStop}%
\bibitem [{\citenamefont {M{\'{e}}nard}\ \emph {et~al.}(2015)\citenamefont
  {M{\'{e}}nard}, \citenamefont {Guissart}, \citenamefont {Brun}, \citenamefont
  {Pons}, \citenamefont {Stolyarov}, \citenamefont {Debontridder},
  \citenamefont {Leclerc}, \citenamefont {Janod}, \citenamefont {Cario},
  \citenamefont {Roditchev}, \citenamefont {Simon},\ and\ \citenamefont
  {Cren}}]{M_nard_2015}%
  \BibitemOpen
  \bibfield  {author} {\bibinfo {author} {\bibfnamefont {G.~C.}\ \bibnamefont
  {M{\'{e}}nard}}, \bibinfo {author} {\bibfnamefont {S.}~\bibnamefont
  {Guissart}}, \bibinfo {author} {\bibfnamefont {C.}~\bibnamefont {Brun}},
  \bibinfo {author} {\bibfnamefont {S.}~\bibnamefont {Pons}}, \bibinfo {author}
  {\bibfnamefont {V.~S.}\ \bibnamefont {Stolyarov}}, \bibinfo {author}
  {\bibfnamefont {F.}~\bibnamefont {Debontridder}}, \bibinfo {author}
  {\bibfnamefont {M.~V.}\ \bibnamefont {Leclerc}}, \bibinfo {author}
  {\bibfnamefont {E.}~\bibnamefont {Janod}}, \bibinfo {author} {\bibfnamefont
  {L.}~\bibnamefont {Cario}}, \bibinfo {author} {\bibfnamefont
  {D.}~\bibnamefont {Roditchev}}, \bibinfo {author} {\bibfnamefont
  {P.}~\bibnamefont {Simon}}, \ and\ \bibinfo {author} {\bibfnamefont
  {T.}~\bibnamefont {Cren}},\ }\href {\doibase 10.1038/nphys3508} {\bibfield
  {journal} {\bibinfo  {journal} {Nat. Phys.}\ }\textbf {\bibinfo {volume}
  {11}},\ \bibinfo {pages} {1013} (\bibinfo {year} {2015})}\BibitemShut
  {NoStop}%
\bibitem [{\citenamefont {Ruby}\ \emph
  {et~al.}(2015{\natexlab{a}})\citenamefont {Ruby}, \citenamefont {Pientka},
  \citenamefont {Peng}, \citenamefont {von Oppen}, \citenamefont {Heinrich},\
  and\ \citenamefont {Franke}}]{Ruby_2015}%
  \BibitemOpen
  \bibfield  {author} {\bibinfo {author} {\bibfnamefont {M.}~\bibnamefont
  {Ruby}}, \bibinfo {author} {\bibfnamefont {F.}~\bibnamefont {Pientka}},
  \bibinfo {author} {\bibfnamefont {Y.}~\bibnamefont {Peng}}, \bibinfo {author}
  {\bibfnamefont {F.}~\bibnamefont {von Oppen}}, \bibinfo {author}
  {\bibfnamefont {B.~W.}\ \bibnamefont {Heinrich}}, \ and\ \bibinfo {author}
  {\bibfnamefont {K.~J.}\ \bibnamefont {Franke}},\ }\href@noop {} {\bibfield
  {journal} {\bibinfo  {journal} {Phys. Rev. Lett.}\ }\textbf {\bibinfo
  {volume} {115}},\ \bibinfo {pages} {087001} (\bibinfo {year}
  {2015}{\natexlab{a}})}\BibitemShut {NoStop}%
\bibitem [{\citenamefont {Hatter}\ \emph {et~al.}(2015)\citenamefont {Hatter},
  \citenamefont {Heinrich}, \citenamefont {Ruby}, \citenamefont {Pascual},\
  and\ \citenamefont {Franke}}]{Hatter2015}%
  \BibitemOpen
  \bibfield  {author} {\bibinfo {author} {\bibfnamefont {N.}~\bibnamefont
  {Hatter}}, \bibinfo {author} {\bibfnamefont {B.~W.}\ \bibnamefont
  {Heinrich}}, \bibinfo {author} {\bibfnamefont {M.}~\bibnamefont {Ruby}},
  \bibinfo {author} {\bibfnamefont {J.~I.}\ \bibnamefont {Pascual}}, \ and\
  \bibinfo {author} {\bibfnamefont {K.~J.}\ \bibnamefont {Franke}},\ }\href
  {http://dx.doi.org/10.1038/ncomms9988} {\bibfield  {journal} {\bibinfo
  {journal} {Nat. Commun.}\ }\textbf {\bibinfo {volume} {6}},\ \bibinfo {pages}
  {8988} (\bibinfo {year} {2015})}\BibitemShut {NoStop}%
\bibitem [{\citenamefont {Ruby}\ \emph {et~al.}(2016)\citenamefont {Ruby},
  \citenamefont {Peng}, \citenamefont {von Oppen}, \citenamefont {Heinrich},\
  and\ \citenamefont {Franke}}]{Ruby_2016}%
  \BibitemOpen
  \bibfield  {author} {\bibinfo {author} {\bibfnamefont {M.}~\bibnamefont
  {Ruby}}, \bibinfo {author} {\bibfnamefont {Y.}~\bibnamefont {Peng}}, \bibinfo
  {author} {\bibfnamefont {F.}~\bibnamefont {von Oppen}}, \bibinfo {author}
  {\bibfnamefont {B.~W.}\ \bibnamefont {Heinrich}}, \ and\ \bibinfo {author}
  {\bibfnamefont {K.~J.}\ \bibnamefont {Franke}},\ }\href@noop {} {\bibfield
  {journal} {\bibinfo  {journal} {Phys. Rev. Lett.}\ }\textbf {\bibinfo
  {volume} {117}},\ \bibinfo {pages} {186801} (\bibinfo {year}
  {2016})}\BibitemShut {NoStop}%
\bibitem [{\citenamefont {{Choi}}\ \emph {et~al.}(2016)\citenamefont {{Choi}},
  \citenamefont {{Rubio-Verd{\'u}}}, \citenamefont {{de Bruijckere}},
  \citenamefont {{Moreno Ugeda}}, \citenamefont {{Lorente}},\ and\
  \citenamefont {{Pascual}}}]{choi_2016}%
  \BibitemOpen
  \bibfield  {author} {\bibinfo {author} {\bibfnamefont {D.-J.}\ \bibnamefont
  {{Choi}}}, \bibinfo {author} {\bibfnamefont {C.}~\bibnamefont
  {{Rubio-Verd{\'u}}}}, \bibinfo {author} {\bibfnamefont {J.}~\bibnamefont {{de
  Bruijckere}}}, \bibinfo {author} {\bibfnamefont {M.}~\bibnamefont {{Moreno
  Ugeda}}}, \bibinfo {author} {\bibfnamefont {N.}~\bibnamefont {{Lorente}}}, \
  and\ \bibinfo {author} {\bibfnamefont {J.~I.}\ \bibnamefont {{Pascual}}},\
  }\href@noop {} {\bibfield  {journal} {\bibinfo  {journal} {arXiv:1608.03752}\
  } (\bibinfo {year} {2016})}\BibitemShut {NoStop}%
\bibitem [{\citenamefont {{Kezilebieke}}\ \emph {et~al.}(2017)\citenamefont
  {{Kezilebieke}}, \citenamefont {{Dvorak}}, \citenamefont {{Ojanen}},\ and\
  \citenamefont {{Liljeroth}}}]{2017arXiv170103288K}%
  \BibitemOpen
  \bibfield  {author} {\bibinfo {author} {\bibfnamefont {S.}~\bibnamefont
  {{Kezilebieke}}}, \bibinfo {author} {\bibfnamefont {M.}~\bibnamefont
  {{Dvorak}}}, \bibinfo {author} {\bibfnamefont {T.}~\bibnamefont {{Ojanen}}},
  \ and\ \bibinfo {author} {\bibfnamefont {P.}~\bibnamefont {{Liljeroth}}},\
  }\href@noop {} {\bibfield  {journal} {\bibinfo  {journal} {arXiv:1701.03288}\
  } (\bibinfo {year} {2017})}\BibitemShut {NoStop}%
\bibitem [{\citenamefont {{Heinrich}}\ \emph {et~al.}(2017)\citenamefont
  {{Heinrich}}, \citenamefont {{Pascual}},\ and\ \citenamefont
  {{Franke}}}]{Franke2017review}%
  \BibitemOpen
  \bibfield  {author} {\bibinfo {author} {\bibfnamefont {B.~W.}\ \bibnamefont
  {{Heinrich}}}, \bibinfo {author} {\bibfnamefont {J.~I.}\ \bibnamefont
  {{Pascual}}}, \ and\ \bibinfo {author} {\bibfnamefont {K.~J.}\ \bibnamefont
  {{Franke}}},\ }\href@noop {} {\bibfield  {journal} {\bibinfo  {journal}
  {arXiv:1705.03672}\ } (\bibinfo {year} {2017})}\BibitemShut {NoStop}%
\bibitem [{\citenamefont {Nadj-Perge}\ \emph {et~al.}(2014)\citenamefont
  {Nadj-Perge}, \citenamefont {Drozdov}, \citenamefont {Li}, \citenamefont
  {Chen}, \citenamefont {Jeon}, \citenamefont {Seo}, \citenamefont {MacDonald},
  \citenamefont {Bernevig},\ and\ \citenamefont {Yazdani}}]{yazdani2014}%
  \BibitemOpen
  \bibfield  {author} {\bibinfo {author} {\bibfnamefont {S.}~\bibnamefont
  {Nadj-Perge}}, \bibinfo {author} {\bibfnamefont {I.~K.}\ \bibnamefont
  {Drozdov}}, \bibinfo {author} {\bibfnamefont {J.}~\bibnamefont {Li}},
  \bibinfo {author} {\bibfnamefont {H.}~\bibnamefont {Chen}}, \bibinfo {author}
  {\bibfnamefont {S.}~\bibnamefont {Jeon}}, \bibinfo {author} {\bibfnamefont
  {J.}~\bibnamefont {Seo}}, \bibinfo {author} {\bibfnamefont {A.~H.}\
  \bibnamefont {MacDonald}}, \bibinfo {author} {\bibfnamefont {B.~A.}\
  \bibnamefont {Bernevig}}, \ and\ \bibinfo {author} {\bibfnamefont
  {A.}~\bibnamefont {Yazdani}},\ }\href {\doibase 10.1126/science.1259327}
  {\bibfield  {journal} {\bibinfo  {journal} {Science}\ }\textbf {\bibinfo
  {volume} {346}},\ \bibinfo {pages} {602} (\bibinfo {year}
  {2014})}\BibitemShut {NoStop}%
\bibitem [{\citenamefont {Ruby}\ \emph
  {et~al.}(2015{\natexlab{b}})\citenamefont {Ruby}, \citenamefont {Pientka},
  \citenamefont {Peng}, \citenamefont {von Oppen}, \citenamefont {Heinrich},\
  and\ \citenamefont {Franke}}]{franke2015}%
  \BibitemOpen
  \bibfield  {author} {\bibinfo {author} {\bibfnamefont {M.}~\bibnamefont
  {Ruby}}, \bibinfo {author} {\bibfnamefont {F.}~\bibnamefont {Pientka}},
  \bibinfo {author} {\bibfnamefont {Y.}~\bibnamefont {Peng}}, \bibinfo {author}
  {\bibfnamefont {F.}~\bibnamefont {von Oppen}}, \bibinfo {author}
  {\bibfnamefont {B.~W.}\ \bibnamefont {Heinrich}}, \ and\ \bibinfo {author}
  {\bibfnamefont {K.~J.}\ \bibnamefont {Franke}},\ }\href {\doibase
  10.1103/PhysRevLett.115.197204} {\bibfield  {journal} {\bibinfo  {journal}
  {Phys. Rev. Lett.}\ }\textbf {\bibinfo {volume} {115}},\ \bibinfo {pages}
  {197204} (\bibinfo {year} {2015}{\natexlab{b}})}\BibitemShut {NoStop}%
\bibitem [{\citenamefont {Pawlak}\ \emph {et~al.}(2016)\citenamefont {Pawlak},
  \citenamefont {Kisiel}, \citenamefont {Klinovaja}, \citenamefont {Meier},
  \citenamefont {Kawai}, \citenamefont {Glatzel}, \citenamefont {Loss},\ and\
  \citenamefont {Meyer}}]{meyer2016}%
  \BibitemOpen
  \bibfield  {author} {\bibinfo {author} {\bibfnamefont {R.}~\bibnamefont
  {Pawlak}}, \bibinfo {author} {\bibfnamefont {M.}~\bibnamefont {Kisiel}},
  \bibinfo {author} {\bibfnamefont {J.}~\bibnamefont {Klinovaja}}, \bibinfo
  {author} {\bibfnamefont {T.}~\bibnamefont {Meier}}, \bibinfo {author}
  {\bibfnamefont {S.}~\bibnamefont {Kawai}}, \bibinfo {author} {\bibfnamefont
  {T.}~\bibnamefont {Glatzel}}, \bibinfo {author} {\bibfnamefont
  {D.}~\bibnamefont {Loss}}, \ and\ \bibinfo {author} {\bibfnamefont
  {E.}~\bibnamefont {Meyer}},\ }\href {\doibase 10.1038/npjqi.2016.35}
  {\bibfield  {journal} {\bibinfo  {journal} {Npj Quantum Information}\
  }\textbf {\bibinfo {volume} {2}},\ \bibinfo {pages} {16035} (\bibinfo {year}
  {2016})}\BibitemShut {NoStop}%
\bibitem [{\citenamefont {Feldman}\ \emph {et~al.}(2017)\citenamefont
  {Feldman}, \citenamefont {Randeria}, \citenamefont {Li}, \citenamefont
  {Jeon}, \citenamefont {Xie}, \citenamefont {Wang}, \citenamefont {Drozdov},
  \citenamefont {Andrei~Bernevig},\ and\ \citenamefont
  {Yazdani}}]{feldman2017}%
  \BibitemOpen
  \bibfield  {author} {\bibinfo {author} {\bibfnamefont {B.~E.}\ \bibnamefont
  {Feldman}}, \bibinfo {author} {\bibfnamefont {M.~T.}\ \bibnamefont
  {Randeria}}, \bibinfo {author} {\bibfnamefont {J.}~\bibnamefont {Li}},
  \bibinfo {author} {\bibfnamefont {S.}~\bibnamefont {Jeon}}, \bibinfo {author}
  {\bibfnamefont {Y.}~\bibnamefont {Xie}}, \bibinfo {author} {\bibfnamefont
  {Z.}~\bibnamefont {Wang}}, \bibinfo {author} {\bibfnamefont {I.~K.}\
  \bibnamefont {Drozdov}}, \bibinfo {author} {\bibfnamefont {B.}~\bibnamefont
  {Andrei~Bernevig}}, \ and\ \bibinfo {author} {\bibfnamefont {A.}~\bibnamefont
  {Yazdani}},\ }\href {http://dx.doi.org/10.1038/nphys3947} {\bibfield
  {journal} {\bibinfo  {journal} {Nat. Phys.}\ }\textbf {\bibinfo {volume}
  {13}},\ \bibinfo {pages} {286} (\bibinfo {year} {2017})}\BibitemShut
  {NoStop}%
\bibitem [{\citenamefont {{Ruby}}\ \emph {et~al.}(2017)\citenamefont {{Ruby}},
  \citenamefont {{Heinrich}}, \citenamefont {{Peng}}, \citenamefont {{von
  Oppen}},\ and\ \citenamefont {{Franke}}}]{franke2017}%
  \BibitemOpen
  \bibfield  {author} {\bibinfo {author} {\bibfnamefont {M.}~\bibnamefont
  {{Ruby}}}, \bibinfo {author} {\bibfnamefont {B.~W.}\ \bibnamefont
  {{Heinrich}}}, \bibinfo {author} {\bibfnamefont {Y.}~\bibnamefont {{Peng}}},
  \bibinfo {author} {\bibfnamefont {F.}~\bibnamefont {{von Oppen}}}, \ and\
  \bibinfo {author} {\bibfnamefont {K.~J.}\ \bibnamefont {{Franke}}},\
  }\href@noop {} {\bibfield  {journal} {\bibinfo  {journal} {arXiv:1704.05756}\
  } (\bibinfo {year} {2017})}\BibitemShut {NoStop}%
\bibitem [{\citenamefont {Kitaev}(2003)}]{Kitaev20032}%
  \BibitemOpen
  \bibfield  {author} {\bibinfo {author} {\bibfnamefont {A.}~\bibnamefont
  {Kitaev}},\ }\href {\doibase http://doi.org/10.1016/S0003-4916(02)00018-0}
  {\bibfield  {journal} {\bibinfo  {journal} {Annals of Physics}\ }\textbf
  {\bibinfo {volume} {303}},\ \bibinfo {pages} {2 } (\bibinfo {year}
  {2003})}\BibitemShut {NoStop}%
\bibitem [{\citenamefont {Nayak}\ \emph {et~al.}(2008)\citenamefont {Nayak},
  \citenamefont {Simon}, \citenamefont {Stern}, \citenamefont {Freedman},\ and\
  \citenamefont {Das~Sarma}}]{Nayak2008}%
  \BibitemOpen
  \bibfield  {author} {\bibinfo {author} {\bibfnamefont {C.}~\bibnamefont
  {Nayak}}, \bibinfo {author} {\bibfnamefont {S.~H.}\ \bibnamefont {Simon}},
  \bibinfo {author} {\bibfnamefont {A.}~\bibnamefont {Stern}}, \bibinfo
  {author} {\bibfnamefont {M.}~\bibnamefont {Freedman}}, \ and\ \bibinfo
  {author} {\bibfnamefont {S.}~\bibnamefont {Das~Sarma}},\ }\href {\doibase
  10.1103/RevModPhys.80.1083} {\bibfield  {journal} {\bibinfo  {journal} {Rev.
  Mod. Phys.}\ }\textbf {\bibinfo {volume} {80}},\ \bibinfo {pages} {1083}
  (\bibinfo {year} {2008})}\BibitemShut {NoStop}%
\bibitem [{\citenamefont {Nadj-Perge}\ \emph {et~al.}(2013)\citenamefont
  {Nadj-Perge}, \citenamefont {Drozdov}, \citenamefont {Bernevig},\ and\
  \citenamefont {Yazdani}}]{Nadj_Perge_2013}%
  \BibitemOpen
  \bibfield  {author} {\bibinfo {author} {\bibfnamefont {S.}~\bibnamefont
  {Nadj-Perge}}, \bibinfo {author} {\bibfnamefont {I.~K.}\ \bibnamefont
  {Drozdov}}, \bibinfo {author} {\bibfnamefont {B.~A.}\ \bibnamefont
  {Bernevig}}, \ and\ \bibinfo {author} {\bibfnamefont {A.}~\bibnamefont
  {Yazdani}},\ }\href {\doibase 10.1103/PhysRevB.88.020407} {\bibfield
  {journal} {\bibinfo  {journal} {Phys. Rev. B}\ }\textbf {\bibinfo {volume}
  {88}},\ \bibinfo {pages} {020407} (\bibinfo {year} {2013})}\BibitemShut
  {NoStop}%
\bibitem [{\citenamefont {Pientka}\ \emph {et~al.}(2013)\citenamefont
  {Pientka}, \citenamefont {Glazman},\ and\ \citenamefont {von
  Oppen}}]{Pientka_2013}%
  \BibitemOpen
  \bibfield  {author} {\bibinfo {author} {\bibfnamefont {F.}~\bibnamefont
  {Pientka}}, \bibinfo {author} {\bibfnamefont {L.~I.}\ \bibnamefont
  {Glazman}}, \ and\ \bibinfo {author} {\bibfnamefont {F.}~\bibnamefont {von
  Oppen}},\ }\href {http://dx.doi.org/10.1103/PhysRevB.88.155420} {\bibfield
  {journal} {\bibinfo  {journal} {Phys. Rev. B}\ }\textbf {\bibinfo {volume}
  {88}},\ \bibinfo {pages} {155420} (\bibinfo {year} {2013})}\BibitemShut
  {NoStop}%
\bibitem [{\citenamefont {Klinovaja}\ \emph {et~al.}(2013)\citenamefont
  {Klinovaja}, \citenamefont {Stano}, \citenamefont {Yazdani},\ and\
  \citenamefont {Loss}}]{Klinovaja2013}%
  \BibitemOpen
  \bibfield  {author} {\bibinfo {author} {\bibfnamefont {J.}~\bibnamefont
  {Klinovaja}}, \bibinfo {author} {\bibfnamefont {P.}~\bibnamefont {Stano}},
  \bibinfo {author} {\bibfnamefont {A.}~\bibnamefont {Yazdani}}, \ and\
  \bibinfo {author} {\bibfnamefont {D.}~\bibnamefont {Loss}},\ }\href {\doibase
  10.1103/PhysRevLett.111.186805} {\bibfield  {journal} {\bibinfo  {journal}
  {Phys. Rev. Lett.}\ }\textbf {\bibinfo {volume} {111}},\ \bibinfo {pages}
  {186805} (\bibinfo {year} {2013})}\BibitemShut {NoStop}%
\bibitem [{\citenamefont {Braunecker}\ and\ \citenamefont
  {Simon}(2013)}]{Braunecker2013}%
  \BibitemOpen
  \bibfield  {author} {\bibinfo {author} {\bibfnamefont {B.}~\bibnamefont
  {Braunecker}}\ and\ \bibinfo {author} {\bibfnamefont {P.}~\bibnamefont
  {Simon}},\ }\href {\doibase 10.1103/PhysRevLett.111.147202} {\bibfield
  {journal} {\bibinfo  {journal} {Phys. Rev. Lett.}\ }\textbf {\bibinfo
  {volume} {111}},\ \bibinfo {pages} {147202} (\bibinfo {year}
  {2013})}\BibitemShut {NoStop}%
\bibitem [{\citenamefont {Kim}\ \emph {et~al.}(2014)\citenamefont {Kim},
  \citenamefont {Cheng}, \citenamefont {Bauer}, \citenamefont {Lutchyn},\ and\
  \citenamefont {Das~Sarma}}]{Kim2014}%
  \BibitemOpen
  \bibfield  {author} {\bibinfo {author} {\bibfnamefont {Y.}~\bibnamefont
  {Kim}}, \bibinfo {author} {\bibfnamefont {M.}~\bibnamefont {Cheng}}, \bibinfo
  {author} {\bibfnamefont {B.}~\bibnamefont {Bauer}}, \bibinfo {author}
  {\bibfnamefont {R.~M.}\ \bibnamefont {Lutchyn}}, \ and\ \bibinfo {author}
  {\bibfnamefont {S.}~\bibnamefont {Das~Sarma}},\ }\href {\doibase
  10.1103/PhysRevB.90.060401} {\bibfield  {journal} {\bibinfo  {journal} {Phys.
  Rev. B}\ }\textbf {\bibinfo {volume} {90}},\ \bibinfo {pages} {060401}
  (\bibinfo {year} {2014})}\BibitemShut {NoStop}%
\bibitem [{\citenamefont {Schecter}\ \emph {et~al.}(2016)\citenamefont
  {Schecter}, \citenamefont {Flensberg}, \citenamefont {Christensen},
  \citenamefont {Andersen},\ and\ \citenamefont {Paaske}}]{Schecter2016}%
  \BibitemOpen
  \bibfield  {author} {\bibinfo {author} {\bibfnamefont {M.}~\bibnamefont
  {Schecter}}, \bibinfo {author} {\bibfnamefont {K.}~\bibnamefont {Flensberg}},
  \bibinfo {author} {\bibfnamefont {M.~H.}\ \bibnamefont {Christensen}},
  \bibinfo {author} {\bibfnamefont {B.~M.}\ \bibnamefont {Andersen}}, \ and\
  \bibinfo {author} {\bibfnamefont {J.}~\bibnamefont {Paaske}},\ }\href
  {\doibase 10.1103/PhysRevB.93.140503} {\bibfield  {journal} {\bibinfo
  {journal} {Phys. Rev. B}\ }\textbf {\bibinfo {volume} {93}},\ \bibinfo
  {pages} {140503} (\bibinfo {year} {2016})}\BibitemShut {NoStop}%
\bibitem [{\citenamefont {Hui}\ \emph {et~al.}(2015)\citenamefont {Hui},
  \citenamefont {Sau},\ and\ \citenamefont {Sarma}}]{Hui_2015_2}%
  \BibitemOpen
  \bibfield  {author} {\bibinfo {author} {\bibfnamefont {H.-Y.}\ \bibnamefont
  {Hui}}, \bibinfo {author} {\bibfnamefont {J.~D.}\ \bibnamefont {Sau}}, \ and\
  \bibinfo {author} {\bibfnamefont {S.~D.}\ \bibnamefont {Sarma}},\ }\href
  {http://dx.doi.org/10.1103/PhysRevB.92.174512} {\bibfield  {journal}
  {\bibinfo  {journal} {Phys. Rev. B}\ }\textbf {\bibinfo {volume} {92}},\
  \bibinfo {pages} {174512} (\bibinfo {year} {2015})}\BibitemShut {NoStop}%
\bibitem [{\citenamefont {Clogston}(1962)}]{Clogston_1962}%
  \BibitemOpen
  \bibfield  {author} {\bibinfo {author} {\bibfnamefont {A.~M.}\ \bibnamefont
  {Clogston}},\ }\href {\doibase 10.1103/physrev.125.439} {\bibfield  {journal}
  {\bibinfo  {journal} {Phys. Rev.}\ }\textbf {\bibinfo {volume} {125}},\
  \bibinfo {pages} {439} (\bibinfo {year} {1962})}\BibitemShut {NoStop}%
\bibitem [{foo()}]{footnote1}%
  \BibitemOpen
  \href@noop {} {}\bibinfo {note} {Eq. (\ref{eq:phase - potential - tan(phi)})
  assumes that the range $\lambda$ of the impurity potential satisfies the
  inequality $k_{\rm F} \lambda \gg 1$. If $\lambda k_{\rm F} \lesssim 1$, a
  correction to Eq.\ (\ref{eq:phase - potential - tan(phi)}) in the form of a
  Cauchy principal value integral has to be included into Eq. (\ref{eq:phase -
  potential - tan(phi)}), see Ref.\ \onlinecite{Clogston_1962}. Since we
  express our results in terms of the phase shifts $\phi_{\sigma}$, these
  corrections are not important for our considerations.}\BibitemShut {Stop}%
\bibitem [{\citenamefont {Pientka}\ \emph {et~al.}(2015)\citenamefont
  {Pientka}, \citenamefont {Peng}, \citenamefont {Glazman},\ and\ \citenamefont
  {von Oppen}}]{Pientka2015}%
  \BibitemOpen
  \bibfield  {author} {\bibinfo {author} {\bibfnamefont {F.}~\bibnamefont
  {Pientka}}, \bibinfo {author} {\bibfnamefont {Y.}~\bibnamefont {Peng}},
  \bibinfo {author} {\bibfnamefont {L.}~\bibnamefont {Glazman}}, \ and\
  \bibinfo {author} {\bibfnamefont {F.}~\bibnamefont {von Oppen}},\ }\href
  {http://stacks.iop.org/1402-4896/2015/i=T164/a=014008} {\bibfield  {journal}
  {\bibinfo  {journal} {Physica Scripta}\ }\textbf {\bibinfo {volume} {2015}},\
  \bibinfo {pages} {014008} (\bibinfo {year} {2015})}\BibitemShut {NoStop}%
\bibitem [{\citenamefont {Pientka}\ \emph {et~al.}(2014)\citenamefont
  {Pientka}, \citenamefont {Glazman},\ and\ \citenamefont {von
  Oppen}}]{Pientka2014}%
  \BibitemOpen
  \bibfield  {author} {\bibinfo {author} {\bibfnamefont {F.}~\bibnamefont
  {Pientka}}, \bibinfo {author} {\bibfnamefont {L.~I.}\ \bibnamefont
  {Glazman}}, \ and\ \bibinfo {author} {\bibfnamefont {F.}~\bibnamefont {von
  Oppen}},\ }\href {\doibase 10.1103/PhysRevB.89.180505} {\bibfield  {journal}
  {\bibinfo  {journal} {Phys. Rev. B}\ }\textbf {\bibinfo {volume} {89}},\
  \bibinfo {pages} {180505} (\bibinfo {year} {2014})}\BibitemShut {NoStop}%
\bibitem [{\citenamefont {Varshalovich}\ \emph {et~al.}(1988)\citenamefont
  {Varshalovich}, \citenamefont {Moskalev},\ and\ \citenamefont
  {Khersonskii}}]{varshalovich1988}%
  \BibitemOpen
  \bibfield  {author} {\bibinfo {author} {\bibfnamefont {D.~A.}\ \bibnamefont
  {Varshalovich}}, \bibinfo {author} {\bibfnamefont {A.~N.}\ \bibnamefont
  {Moskalev}}, \ and\ \bibinfo {author} {\bibfnamefont {V.~K.}\ \bibnamefont
  {Khersonskii}},\ }\href@noop {} {\emph {\bibinfo {title} {Quantum theory of
  angular momentum}}}\ (\bibinfo  {publisher} {World scientific},\ \bibinfo
  {year} {1988})\BibitemShut {NoStop}%
\bibitem [{\citenamefont {Lang}\ and\ \citenamefont {Schwab}(2015)}]{lang2015}%
  \BibitemOpen
  \bibfield  {author} {\bibinfo {author} {\bibfnamefont {A.}~\bibnamefont
  {Lang}}\ and\ \bibinfo {author} {\bibfnamefont {C.}~\bibnamefont {Schwab}},\
  }\href {http://dx.doi.org/10.1214/14-AAP1067} {\bibfield  {journal} {\bibinfo
   {journal} {Ann. Appl. Probab.}\ }\textbf {\bibinfo {volume} {25}},\ \bibinfo
  {pages} {3047} (\bibinfo {year} {2015})}\BibitemShut {NoStop}%
\end{thebibliography}%

\end{document}